\documentclass[onecolumn]{aa520}
\usepackage{graphicx,amsmath,array,float,ifthen}
\usepackage{txfonts}
\newcommand{\Point}{{\bf P}}
\newcommand{\Noise}{{\bf N}}
\def\arcs{\ifmmode {^{\scriptscriptstyle\prime\prime}}
          \else $^{\scriptscriptstyle\prime\prime}$\fi}
\def\arcm{\ifmmode {^{\scriptscriptstyle\prime}}
          \else $^{\scriptscriptstyle\prime}$\fi}
\newdimen\sa  \newdimen\sb
\def\parcs{\sa=.07em \sb=.03em
     \ifmmode $\rlap{.}$^{\scriptscriptstyle\prime\kern -\sb\prime}$\kern -\sa$
     \else \rlap{.}$^{\scriptscriptstyle\prime\kern -\sb\prime}$\kern -\sa\fi}
\def\parcm{\sa=.08em \sb=.03em
     \ifmmode $\rlap{.}\kern\sa$^{\scriptscriptstyle\prime}$\kern-\sb$
     \else \rlap{.}\kern\sa$^{\scriptscriptstyle\prime}$\kern-\sb\fi}

\begin{document}
   \title{Making Maps from Planck LFI 30GHz Data}


   \author{%
  M.\ A.\ J.\ Ashdown\inst{1,2}
  \and
  C.\ Baccigalupi\inst{3,4}
  \and
  A.\ Balbi\inst{5}
  \and
  J.\ G.\ Bartlett\inst{6}
  \and
  J.\ Borrill\inst{7,8}
  \and
  C.\ Cantalupo\inst{8,7}
  \and
  G.\ de Gasperis\inst{5}
  \and
  K.\ M.\ G\'{o}rski\inst{9,10,11}
  \and
  V.\ Heikkil\"{a}\inst{12}
  \and
  E.\ Hivon\inst{10,14}
  \and
  E.\ Keih\"{a}nen\inst{12,13}
  \and
  H.\ Kurki-Suonio\inst{12}
  \and
  C.\ R.\ Lawrence\inst{9}
  \and
  P.\ Natoli\inst{5}
  \and
  T.\ Poutanen\inst{12,13}
  \and
  S.\ Prunet\inst{14}
  \and
  M.\ Reinecke\inst{15}
  \and
  R.\ Stompor\inst{7,8,6}
  \and
  B.\ Wandelt\inst{16,17}\\
  (The Planck CTP Working Group)
     }

   \offprints{H.~Kurki-Suonio, \email{hannu.kurki-suonio@helsinki.fi} }

   \institute{%
  Astrophysics Group, Cavendish Laboratory, J J Thomson Avenue,
  Cambridge CB3 0HE, United Kingdom.
  \and
  Institute of Astronomy, Madingley Road, Cambridge CB3 0HA,
  United Kingdom.
  \and
  Institut f\"{u}r Theoretische Astrophysik,
  Universit\"{a}t Heidelberg, Albert-\"{U}berle-Str. 2,
  D-69120, Heidelberg, Germany.
  \and
  SISSA/ISAS, Via Beirut 4, I-34014 Trieste, and INFN,
  Sezione di Trieste, Via Valerio 2, I-34127, Italy
  \and
  Dipartimento di Fisica, Universit\`{a} di Roma ``Tor Vergata'',
  via della Ricerca Scientifica 1, I-00133 Roma, Italy.
  \and
  Laboratoire Astroparticule \& Cosmologie,
  11 place Marcelin Berthelot, 75231 Paris Cedex 05, France
  (UMR 7164 CNRS, Universit\'e Paris 7, CEA, Observatoire de Paris).
  \and
  Computational Research Division, Lawrence Berkeley National
  Laboratory, Berkeley CA 94720, U.\ S.\ A.
  \and
  Space Sciences Laboratory,
  University of California Berkeley, Berkeley CA 94720, U.\ S.\ A.
  \and
  Jet Propulsion Laboratory, California Institute of Technology, 4800 Oak
  Grove Drive, Pasadena CA 91109, U.\ S.\ A.
  \and
  California Institute of Technology, Pasadena CA 91125, U.\ S.\ A.
  \and
  Warsaw University Observatory, Aleje Ujazdowskie 4, 00478 Warszawa, Poland.
  \and
  University of Helsinki, Department of Physical Sciences,
  P. O. Box 64, FIN-00014 Helsinki, Finland.
  \and
  Helsinki Institute of Physics, P.\ O.\ Box 64, FIN-00014 Helsinki,
  Finland.
  \and
  Institut d'Astrophysique de Paris, 98 bis Boulevard Arago,
  F-75014 Paris, France.
  \and
  Max-Planck-Institut f\"{u}r Astrophysik, Karl-Schwarzschild-Str.~1,
  D-85741 Garching, Germany.
  \and
  Department of Physics, University of Illinois at
  Urbana-Champaign, 1110 West Green Street, Urbana IL 61801, U.\ S.\ A.
  \and
  Department of Astronomy, University of Illinois at
  Urbana-Champaign, 1002 West Green Street, Urbana IL 61801, U.\ S.\ A.
     }

\date{Received date / Accepted date}

   \abstract{This paper is one of a series describing the performance
   and accuracy of map-making codes as assessed by the {\sc Planck} CTP working
   group. We compare the performance of multiple codes written by
   different groups for making polarized maps from {\sc Planck}-sized,
   all-sky cosmic microwave background (CMB) data. Three of the
   codes are based on destriping algorithm, whereas the other three
   are implementations of a maximum-likelihood algorithm. Previous
   papers in the series described simulations at 100~GHz (Poutanen
   et al.~\cite{poutanen06}) and 217~GHz (Ashdown et al.~\cite{ashdown06}). In this paper we
   make maps (temperature and polarisation) from the simulated
   one-year observations of four 30~GHz detectors of {\sc Planck} Low
   Frequency Instrument (LFI). We used {\sc Planck} Level S simulation
   pipeline to produce the observed time-ordered-data streams (TOD).
   Our previous studies considered polarisation observations for
   the CMB only. For this paper we increased the realism of the
   simulations and included polarized galactic foregrounds to our
   sky model. Our simulated TODs comprised of dipole, CMB, diffuse
   galactic emissions, extragalactic radio sources, and detector
   noise. The strong subpixel signal gradients arising from the
   foreground signals couple to the output map through the map-making
   and cause an error (signal error) in the maps. Destriping codes
   have smaller signal error than the maximum-likelihood codes. We
   examined a number of schemes to reduce this error. On the other
   hand, the maximum-likelihood map-making codes can produce maps
   with lower residual noise than destriping codes.
   }

   \maketitle

\section{Introduction}

{\sc Planck} is an ESA/NASA mission to measure the anisotropy of the
temperature and polarization of the cosmic microwave background
(CMB) radiation over the whole sky to unprecedented accuracy at high
angular resolution, and at the largest number of frequency channels
employed by a single CMB experiment up to date. The mission goals
are ambitious and the corresponding demands on efficiency and
accuracy of the associated data analysis are quite extreme. The
first essential stage of data analysis involves processing of
time-ordered data (TOD) and production of  sky maps at each
frequency band of the experiment. This task is nontrivial because
temporal correlations of the detector noise streams due to
$1/f$-spectrum noise can lead to artifacts (e.g. stripes) in the sky
maps. A number of CMB map-making techniques have been developed for
the purpose of efficient suppression of this adverse effect.

This paper is one of a series describing the performance and
accuracy of map-making algorithms/codes as assessed by the {\sc
Planck} CTP working group. Previous papers in the series described
simulations at 100~GHz (Poutanen et al. \cite{poutanen06}) and
217~GHz (Ashdown et al. \cite{ashdown06}). These simulations
employed a number of simplifications. Poutanen et al.
\cite{poutanen06} considered only temperature observations (no
polarization). Ashdown et al. \cite{ashdown06} assumed a simplified
model of instantaneous measurement on the sky (i.e., no integration
for non-zero intervals as detectors scanned across the sky or
bolometer time constant were accounted for), symmetric beams for the
detectors, and a simplified sky model with only CMB temperature and
polarization included. Both studies assumed no gaps in the data
streams (i.e., we assumed a continuous, uninterrupted data streams).

In this paper we increase the realism of the simulations by
expanding the sky model to include polarized galactic and
extragalactic foregrounds. This choice was based on our previous
work (Poutanen et al. \cite{poutanen06}; Ashdown et al.
\cite{ashdown06}). It turned out that the generation of error (e.g.
stripes) in maps is sensitive to the presence of sharp signal
gradients in the observed sky on small angular scales, and the
interplay of this with the pixelization scales of the output sky
maps.

For the simulations described in this paper, we change from 217\,GHz
to 30\,GHz, for practical reasons: 30\,GHz has the lowest resolution
(FWHM$\approx$32\arcm, FWHM = full width half maximum), the smallest
number of optical elements (two) and detectors (four), and the
lowest sampling rate of all the {\sc Planck} frequencies.  This
lessens the data volume and consequent computational demands for
complete map making at this frequency, which in turn increases the
rate of repetitive analysis given finite computing resources, while
retaining all qualitative features required for meaningful study of
residual striping due to correlated noise and scanning.

Moreover, galactic foreground emission is very strong at 30\,GHz,
and its effect on map-making due to the presence of sharp galactic
signal gradients is expected to be nearly extreme amongst {\sc
Planck} frequency channels.  Hence, our exercise in this paper is
well focused on the ability of map-making algorithms to handle
strong galactic foregrounds without resulting loss of quality in the
derived sky maps.

In Sect.~3 we discuss in detail the production of the simulated data
used in this study, since such details seem not to have been
published elsewhere.  In Sect.~3.7 we verify the accuracy of the
simulated data.  In Sect.~4 we discuss the results of the map-making
comparison.  We discuss at length the effect of sub-pixel structure
in the data on the quality of the output maps, since this was a new
level of added realism in the simulated data not present in our
previous study (Ashdown et al.~~\cite{ashdown06}).

\section{Codes}

The map-making codes compared in this paper are described in Ashdown
et al.~\cite{ashdown06}.  They comprise three ``destriping" codes
(Polar, Springtide, and Madam), and three ``optimal" or generalized
least squares (GLS) codes (MapCUMBA, MADmap, and ROMA). These codes
and the computing resources required by them are thoroughly
discussed in Ashdown et al.~\cite{ashdown06}.

\def\deg{\ifmmode ^{\circ}
         \else $^{\circ}$\fi}
\def\pdeg{\ifmmode $\setbox0=\hbox{$^{\circ}$}\rlap{\hskip.11\wd0 .}$^{\circ}
          \else \setbox0=\hbox{$^{\circ}$}\rlap{\hskip.11\wd0 .}$^{\circ}$\fi}
\def\arcs{\ifmmode {^{\scriptscriptstyle\prime\prime}}
          \else $^{\scriptscriptstyle\prime\prime}$\fi}
\def\arcm{\ifmmode {^{\scriptscriptstyle\prime}}
          \else $^{\scriptscriptstyle\prime}$\fi}
\newdimen\sa  \newdimen\sb
\def\parcs{\sa=.07em \sb=.03em
     \ifmmode $\rlap{.}$^{\scriptscriptstyle\prime\kern -\sb\prime}$\kern -\sa$
     \else \rlap{.}$^{\scriptscriptstyle\prime\kern -\sb\prime}$\kern -\sa\fi}
\def\parcm{\sa=.08em \sb=.03em
     \ifmmode $\rlap{.}\kern\sa$^{\scriptscriptstyle\prime}$\kern-\sb$
     \else \rlap{.}\kern\sa$^{\scriptscriptstyle\prime}$\kern-\sb\fi}

\section{Inputs} 
 \label{sec:inputs}

\subsection{Scan strategy} \label{subsec:scanning}

The TOD used as inputs in the map-making were generated by the {\sc
Planck} Level-S software (Reinecke et al.~\cite{reinecke06}).  The
correspondence between the sample sequence of the TOD and locations
on the sky is determined by the scan strategy.  The {\sc Planck}
satellite will orbit the second Lagrangian point ($L_2$) of the
Earth-Sun system (Dupac \& Tauber~\cite{dupac05}), where it will
stay near the ecliptic plane and the Sun-Earth line.

{\sc Planck} will spin at $\sim1$\,rpm on an axis pointed near the
Sun-Earth line. The angle between the spin axis and the optical axis
of the telescope is $85\degr$; the detectors will scan nearly great
circles on the sky.  The spin axis is repointed hourly, remaining
fixed between repointings.  Different scan strategies considered for
{\sc Planck} (Dupac \& Tauber~\cite{dupac05}) differ in the path on
the sky followed by the spin axis. We used a ``cycloidal" scan
strategy, in which the spin axis follows a circular path around the
anti-Sun direction with a period of six months, and the angle
between the spin axis and the anti-Sun direction is 7\pdeg5. This is
the minimum angle that results in all feeds covering the entire sky.
We assumed a non-ideal satellite motion, with spin axis nutation and
variations in the satellite spin rate.  The nutation amplitude and
the deviation from the nominal spin rate were chosen randomly at
every repointing from a truncated Gaussian probability distribution
with parameters (0\parcm5 rms, 2\arcm\ max) for the nutation
amplitude and (0\pdeg12\,s$^{-1}$ rms, 0\pdeg3\,s$^{-1}$ max) for
the spin rate deviation. The abbreviation ``rms" refers to the
root-mean-square.

TODs 366\,days long were generated for the four 30\,GHz LFI detectors, with
$1.028\times 10^9$ samples per detector corresponding to a sampling frequency of
$f_\mathrm{s}=32.5$\,Hz.  TODs for CMB (C), dipole (D), foreground (F), and
instrument noise (N) were generated individually for every detector and were
stored in separate files.  The CMB and foreground TODs contained the effects of
both temperature and polarisation anisotropies.  Maps were later made from
different combinations of these four TODs.

We used the HEALPix\footnote{http://healpix.jpl.nasa.gov} pixelisation scheme
(G\'orski et al.~\cite{gorski99}, \cite{gorski05a}) with $N_{\rm side}=512$.  A
map of the full sky contains $12N_{\rm side}^2$ pixels.  The Stokes parameters Q
and U at a point on the sky are defined in a reference coordinate system
($\mathbf{e}_{\theta},\mathbf{e}_{\varphi},\mathbf{n}$), where the
unit vector $\mathbf{e}_{\theta}$ is along the increasing $\theta$ direction,
$\mathbf{e}_{\varphi}$ is along the increasing $\varphi$ direction, and
$\mathbf{n}$ points to the sky (G\'orski et al.~\cite{gorski05b}). The angles
$\theta$ and $\varphi$ are the polar and azimuth angles of the spherical polar
coordinate system used for the celestial sphere.

\begin{figure} [!ht]
    \begin{center}
    \includegraphics[scale=0.33,angle=90]{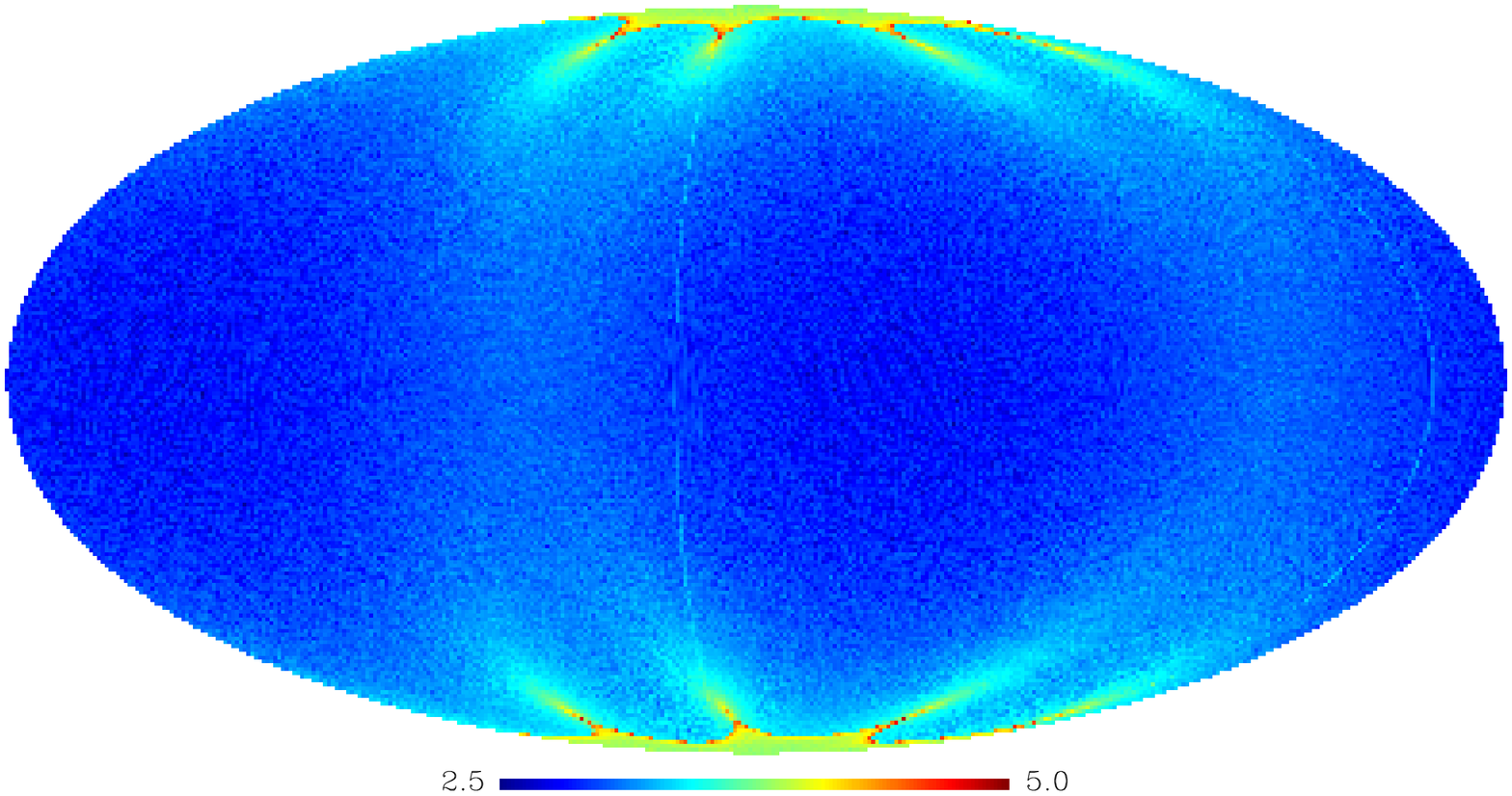}
    \includegraphics[scale=0.33,angle=90]{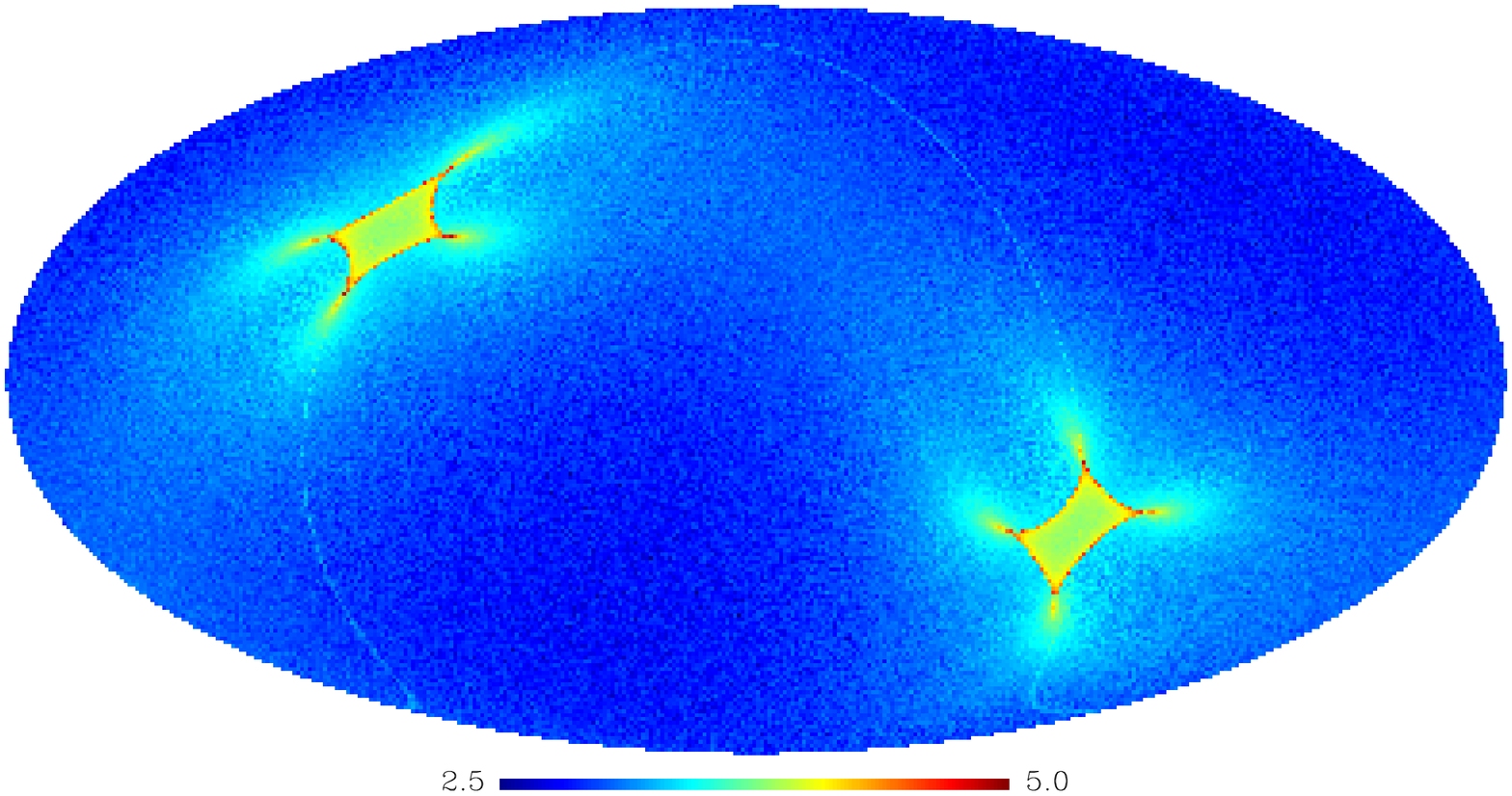}
    \end{center}
  \caption{Number of hits per pixel ($n_\mathrm{hit}$) for the
scan strategy applied in this study. The hit map is shown in the
ecliptic (left) and galactic (right) coordinates. The latter map
shows the areas of the ecliptic poles more clearly. Both maps
include the hits of all four LFI 30~GHz detectors. The scale is
$\log_{10}(n_{\rm{hit}})$. (The original version of this paper with
better-quality figures is available at
www.helsinki.fi/\~{}tfo\_cosm/tfo\_planck.html .)}
  \label{fig:hits}
 \end{figure}

The number of hits per pixel from all detectors is shown in
Fig.~\ref{fig:hits}.  At this resolution every pixel was hit.

\subsection{Telescope beams} \label{subsec:beams}

We assumed identical, circularly symmetric Gaussian telescope beams
for every detector. The FWHM of the beams was 32\parcm19. There are
two 30\,GHz feed horns (with two detectors corresponding to two
polarisation directions per horn). A coordinate system is defined
for each detector, with $z$-axis along the direction of the beam
center and $x$ and $y$-axes perpendicular to the pointing, called
the {\it main beam coordinate system\/}. The spin 0 and $\pm$2
spherical harmonic coefficients ($b_{\ell m}$ and $_{\pm2}b_{\ell
m}$) of the beam response at the reference pointing and orientation
were generated for multipoles up to $\ell_{\rm max}$ = 3000
(Challinor et al.~\cite{challinor00}; Reinecke et
al.~\cite{reinecke06}). The T, E, and B mode coefficients of the
beam were obtained as $b_{\ell m}^{\rm T} = b_{\ell m}$, $b_{\ell
m}^{\rm E} = -(_{+2}b_{\ell m} + _{-2}b_{\ell m})/2$ and $b_{\ell
m}^{\rm B} = i(_{+2} b_{\ell m} - _{-2}b_{\ell m})/2$.  The
coefficients $b_{\ell m}^{\rm T}$, $b_{\ell m}^{\rm E}$ and $b_{\ell
m}^{\rm B}$ are called the {\it beam $b_{\ell m}$} in this study.

\subsection{Noise TOD} \label{subsec:noise}

The instrument noise was a sum of white and correlated $1/f$ noise.
The power spectral density (PSD) of the noise was
 \begin{equation}
   P(f) = \left[1+\left(\frac{f_{\rm k}}{f}\right)^{\alpha}\right]\frac{\sigma^2}{f_\mathrm{s}},
   \quad(f>f_\mathrm{min}), \label{psd}
 \end{equation}
where $f_{\rm k}$ is the knee frequency, i.e., the frequency at
which the $1/f$ and white noise are equal, $f_s$ is the sampling
frequency, and $\sigma$ is the nominal white noise standard
deviation per sample integration time ($t = 1/f_{\mathrm{s}}$).
Below $f_{\rm min}$ the noise spectrum becomes flat. (A $1/f$
spectrum extending to $f=0$ is unphysical.)  The spectral slope of
the correlated part of the noise is given by $\alpha$. The values of
the noise parameters used in this study were: $f_{\rm k}$ = 0.05~Hz,
$f_{\rm min} = 1.15 \times 10^{-5}$~Hz, $\sigma$ = 1350~$\mu$K (CMB
scale) and $\alpha$ = 1.7. They represent realistic expected noise
performance of the instrument. We used the stochastic differential
equation (SDE) algorithm (from Level-S) to generate the TODs of the
instrumental noise.  The noise samples were generated in 6-day
chunks with no correlation between the chunks.  No correlation was
assumed between the noise TODs of different detectors.  For the GLS
and Madam map-making codes, perfect knowledge of the above noise
parameter values was assumed in the map-making phase.

\subsection{Dipole} \label{subsec:dipole}

The temperature Doppler shift that arises from the (constant) motion of the
solar system relative to the last scattering surface was included. The Doppler
shift arising from the satellite motion relative to the Sun was not
included.

\subsection{CMB and foregrounds} \label{subsec:cmb_fg}

The CMB and foreground emissions of the sky were modelled with the
sets of spherical harmonic coefficients $a_{C,\ell m}^{\rm T,E,B}$
and $a_{F,\ell m}^{\rm T,E,B}$.  Here C (F) refers to the CMB
(foreground), T refers to the temperature, and E and B refer to the
polarisation modes.  These expansion coefficients are called
{\it sky $a_{\ell m}$} in this study. The sky $a_{\ell m}$ were
determined for multipoles up to $\ell_{\rm max}$ = 3000.

The expansion coefficients obtained by convolving the sky $a_{\ell m}$ with the
beam $b_{\ell m}$ are called the {\it input $a_{\ell m}$}. The map (for the
Stokes parameters I, Q and U) made from the input $a_{\ell m}$ is called the
{\it input map} in this study.

In this section we describe how the CMB and foreground signals were
simulated for 30\,GHz\footnote{CMB and foreground templates are a
subset of the emissions included in the version 0.1 of the Planck
reference sky that is available in www.planck.fr/heading79.html.}.
In this work the foreground signal is the sum of unresolved
extragalactic components and diffuse emission from the Galaxy.
These signals are represented as template maps with
1\parcm7 pixel size ($N_{\rm side}=2048$). The
coefficients $a_{F,\ell m}^{\rm T,E,B}$ were determined from the
foreground template map using the anafast code of the HEALPix
package.

\subsubsection{CMB}
   \label{subsubsec:cmb_template}

The CMB pattern is the same that we used in our earlier study
(Ashdown et al. \cite{ashdown06}).

For the total intensity at $\ell \le 70$, the $a_{C,\ell m}^{\rm T}$
coefficients were taken directly from the output of the anafast code
running on the  WMAP internal linear combination (ILC) CMB template
(Bennett et al.~\cite{bennett03b})\footnote{The WMAP ILC template is
available in
lambda.gsfc.nasa.gov/product/map/dr1/m$_{-}$products.cfm.}.
 The coefficients of the E mode polarization,
$a_{C,\ell m}^{\rm E}$, at $\ell \le 70$, were obtained as
 \begin{equation}
 a_{C,\ell m}^{\rm E} =
       a_{C,\ell m}^{\rm T}\cdot\frac{C_{\ell}^{\rm TE}}{C_{\ell}^{\rm TT}}
     + \left(\frac{x_{\ell m} + iy_{\ell m}}{\sqrt{2}}\right)\cdot\sqrt{C_{\ell}^{\rm EE}
     - \frac{C_{\ell}^{\rm TE}}{C_{\ell}^{\rm TT}}C_{\ell}^{\rm TE}},
 \label{cmb_e_mode}
 \end{equation}
where $C_{\ell}^{\rm XY}$ (X,Y = T,E) is the best fit angular power
spectrum to the WMAP, ACBAR, and CBI data\footnote{$C_{\ell}^{\rm
XY}$ was determined by the WMAP team and it is available in
lambda.gsfc.nasa.gov/product/map/dr1/lcdm.cfm}. The quantities
$x_{\ell m}$ and $y_{\ell m}$ are Gaussian distributed random
variables with zero mean and unit variance.  The imaginary part
($y_{\ell m}$) and the $\sqrt{2}$-factor were not used for $m = 0$.

At small angular scales ($\ell > 70$), the $a_{C,\ell m}^{\rm T,E}$ coefficients
were a random realization of $C_{\ell}^{\rm XY}$ (using the spectrum-to-alm mode
of the synfast code of the HEALPix package).

More accurate representations of the CMB are available (e.g.,
O'Dwyer et al. \cite{odwyer04}), but small differences in the input
sky are unimportant for our purposes.

\subsubsection{Extragalactic emission} \label{subsubsec:egs}

At 30\,GHz, the dominant contributions are given by the radio sources. They
have been simulated according to the recent data in total intensity and
polarization reaching 20\,GHz (see \cite{tucci04} and references therein).  The
information extracted from those observations is entirely statistical.  In
particular, the flux number counts as a function of frequency, and the
distribution of the polarization degree are inferred from the observations.
Two populations are considered, namely flat and steep spectrum sources, with
total intensity spectral index of $-2$ and $-1.2$, respectively%
\footnote{Measurements of sources from the ATCA 20\,GHz Pilot Survey
show that the sources that will affect CMB observations have spectra
that are not well-characterized by a simple spectral index (Sadler
et al.~2006).  Future simulations should take this into account.}.
From the polarization distribution, a polarization degree of 2.7\%
and 4.8\% is adopted to the flat and steep spectrum species,
respectively.  Finally, the polarization angle is assigned randomly
to each source.

\subsubsection{Diffuse Galactic emission} \label{subsubsec:dge}

At 30\,GHz, the relevant emission processes from our own galaxy are synchrotron
emission from free electrons spiraling around the Galactic magnetic field, and
bremsstrahlung emitted by electrons scattering off hydrogen ions.  Thermal
emission from dust grains dominates galactic emission above about 70\,GHz; we
include it here because peaks of dust emission in the galactic plane are
relevant even at 30\,GHz.

For synchrotron emission we use the model of \cite{giardino02}.  On
scales of one degree or larger, the 408\,MHz sky (\cite{haslam82})
is scaled point by point with a spectral index determined from
408\,MHz and 1420\,MHz data (\cite{Reich}).  Structure is extended
to smaller angular scales assuming a power law power spectrum
matched in amplitude to the local 408\,MHz emission, with
$C_\ell\sim\ell^{-2}$.  For polarization we assume the theoretical
maximum value for synchrotron emission of $75\%$, and take the
polarization direction from the observations at low and medium
galactic latitudes.  These measures give a rather high fluctuation
level reflecting the small scale structure of the galactic magnetic
field (\cite{uyaniker99}; Duncan et al. \cite{duncan99}), scaling as
$\ell^{-2}$ from sub-degree to arcminute scales (\cite{tucci02}),
and consistent with recent observations at medium galactic latitudes
(\cite{carretti05}). The template for the polarization angle was
obtained by adopting the form above for the angular power spectrum,
and assuming a Gaussian distribution.

The intensity of the thermal dust emission is well known at
100\,$\mu$m, and can be extrapolated to microwave frequencies using
the emissivity and temperature of two thermal components
(\cite{finkbeiner99}).  We used model~8 of \cite{finkbeiner99}, with
emissivities and temperatures fixed across the sky. The polarized
emission from the diffuse thermal dust has been detected for the
first time in the Archeops data (Beno\^{\i}t et al.
\cite{benoit04}), which showed a fractional polarization of 5\%. The
pattern of the polarization angle is much less certain. It is caused
by the magnetized dust grains, which are aligned along the galactic
magnetic field (\cite{prunet98}).  Since the geometry and
composition of the dust grains are still very uncertain, the
simplest assumption is that the galactic magnetic field is 100\%
efficient in imprinting the polarization angle pattern to the
synchrotron and dust emissions (Baccigalupi \cite{baccigalupi03}).

\subsection{Convolution of sky and beam for the TOD} \label{subsec:convolution}

To generate the CMB and foreground TOD, we need to convolve the CMB
and foreground skies with the telescope beam for every sample of the
\hbox{TOD}. We used the total convolution algorithm originally
introduced by Wandelt \& G\'orski~\cite{wandelt01} to treat the
temperature anisotropies, and generalized for polarisation by
Challinor et al.~\cite{challinor00}. The polarized total convolution
algorithm has been implemented as a part of Level-S (Reinecke et
al.~\cite{reinecke06}).

Specifically, the i$^{\rm th}$ sample of a detector is represented as
 \begin{equation}
    s_i = \sum_{m'' = -m_{\rm max}}^{m_{\rm max}}\sum_{m,m' =
         -\ell_{\rm max}}^{\ell_{\rm max}}{T_{mm'm''}e^{i(m\varphi_i +
             m'\theta_i + m''\psi_i)}},
 \label{tod_sample}
 \end{equation}
where ($\theta_i,\varphi_i$) are the spherical coordinates of the
pointing of the i$^{\rm th}$ sample. The angle from the
$\mathbf{e}_{\theta}$ axis to the x-axis of the main beam coordinate
system is $\psi_i$. The largest multipole of the sky $a_{\ell m}$
and the beam $b_{\ell m}$ is $\ell_{\rm max}$ ($\ell_{\rm max}$ =
3000 in this study). The largest $m''$ (of the beam $b_{\ell m}$)
that is required to describe the beam response is $m_{\rm max}$. For
the circularly symmetric beam the $m''$ values $-2$, 0, and 2 are
sufficient to fully describe the convolution (Challinor et
al.~\cite{challinor00}).  The quantity $T_{mm'm''}$ represents the
convolution of the sky with the beam (Reinecke et
al.~\cite{reinecke06}) and it is determined from the three sets of
parameters: sky $a_{\ell m}$'s, beam $b_{\ell m}$'s and an angle
$\psi_{\rm pol}$, which is the angle between the x-axis of the main
beam coordinate system and the polarisation sensitive direction of
the detector. In the i$^{\rm th}$ sample the angle from the
$\mathbf{e}_{\theta}$ axis to the polarisation sensitive direction
of the detector is $\psi_i + \psi_{\rm pol}$.

The calculation of the samples $s_i$ in their true pointings would
be an unrealistically tedious task for the typical mission times (12~months in
this study). Therefore in the total convolution algorithm the $m$ and $m'$ sums
of Eq.~(\ref{tod_sample}) are first carried out using a 2-dimensional fast
Fourier transform (FFT), and the results are tabulated in 2-dimensional equally
spaced grids $T_{m''}(\theta,\varphi)$, where ($\theta,\varphi$) tabulates the
grid intersections.  The size of the grid is $[0,\pi]$ in the $\theta$
direction and $[0,2\pi]$ in the $\varphi$ direction.  The grid spacing is
$\pi/\ell_{\rm max}$ in both directions.  For the circularly symmetric beams
there are three grids (for $m'' = -2$, 0, and 2). Because the samples $s_i$
are real, $T_{0}(\theta,\varphi)$ is real, and $T_{-2}(\theta,\varphi)$ =
$T^{\ast}_{2}(\theta,\varphi)$.
For the case of the circularly symmetric beam, the Level-S total
convolver software stores three grids: $T_{0}(\theta,\varphi)$ and
the real and imaginary parts of $T_{2}(\theta,\varphi)$ (Reinecke et
al.~\cite{reinecke06}). These grids are called the {\it ring sets\/}
in Level-S.

Level-S uses polynomial interpolation to determine the ring set value
at a given pointing from the tabulated values.  After the interpolation the
final TOD sample value is calculated as
 \begin{equation}
   \tilde{s}_i = \widetilde{T}_{-2}(\theta_i,\varphi_i)e^{-i2\psi_i} +
   \widetilde{T}_{0}(\theta_i,\varphi_i) +
   \widetilde{T}_{2}(\theta_i,\varphi_i)e^{i2\psi_i}.
 \label{tod_estimate1}
 \end{equation}
The symbol $\tilde{s}_i$ is used for the calculated TOD sample to
indicate that it is an approximation (due to the interpolation) of
the true sample value $s_i$ (see Eq.~(\ref{tod_sample})). In
Eq.~(\ref{tod_estimate1}), $\widetilde{T}_{m''}(\theta_i,\varphi_i)$
is the ring set value interpolated from the tabulated ring set
values. The relation between the ring set values and the Stokes
parameters is
 \begin{equation}
   \widetilde{T}_{0}(\theta_i,\varphi_i) =
   \widetilde{I}(\theta_i,\varphi_i) \quad \quad
   \widetilde{T}_{2}(\theta_i,\varphi_i) =
   \frac{1}{2}\left(\widetilde{Q}(\theta_i,\varphi_i) -
   i\widetilde{U}(\theta_i,\varphi_i)\right)e^{i2\psi_{\rm pol}}
 \label{stokes_ringsets}
 \end{equation}
Inserting these into Eq. (\ref{tod_estimate1}) leads to the standard
formula for the TOD sample
 \begin{equation}
   \tilde{s}_i = \widetilde{I}(\theta_i,\varphi_i) +
   \widetilde{Q}(\theta_i,\varphi_i)\cos{[2(\psi_i + \psi_{\rm pol})]} +
   \widetilde{U}(\theta_i,\varphi_i)\sin{[2(\psi_i + \psi_{\rm pol})]}.
 \label{TOD_estimate2}
 \end{equation}

The interpolation error causes $\widetilde{I}$, $\widetilde{Q}$ and
$\widetilde{U}$ to deviate from the Stokes parameters of the input
map.  It was discovered in earlier studies that a high-order
interpolation is required to make this error sufficiently small.
Accordingly, we used 11$^{\rm th}$ order polynomial interpolation
when producing the CMB and foreground TODs for this study. The
interpolation error that we make is demonstrated in
Sect.~\ref{sec:verification}.

\subsection{Verification} \label{sec:verification}

To assess the effects that are present in the (noiseless) CMB and foreground
TODs, we compared three CMB maps and three foreground maps. The maps were the
input map, ring set map and the binned noiseless map. We used two map
resolutions for the comparisons, $N_{\rm side} = 512$ and $N_{\rm side} = 1024$.

The {\em input maps} were discussed in Sect.~\ref{subsec:cmb_fg}.

The {\it ring set map\/} is obtained by interpolating the ring set
values ($\widetilde{T}_{0}(\theta_i,\varphi_i)$ and
$\widetilde{T}_{2}(\theta_i,\varphi_i)$) of the HEALPix pixel
centers from the tabulated ring set values and then solving the
corresponding Stokes parameters $\widetilde{I}(\theta_i,\varphi_i)$,
$\widetilde{Q}(\theta_i,\varphi_i)$ and
$\widetilde{U}(\theta_i,\varphi_i)$ (see
Eq.~(\ref{stokes_ringsets})).  The pixel triplets ($\widetilde{I}$,
$\widetilde{Q}$, and $\widetilde{U}$) made the ring set map. It is
expected that the difference between the ring set map and the input
map is mainly caused by the interpolation error.  As in the TOD
generation, we used 11$^{\rm th}$ order polynomial interpolation to
produce the ring set maps.  The ring set maps of this study were
made from the ring sets of the detector LFI-27a.  It was verified
that the ring sets of the other detectors produced identical ring
set maps.

The {\it binned noiseless map\/} is obtained by binning the TOD samples in
the map pixels
 \begin{equation}
 \mathbf{m}^{\rm B} = (\mathbf{P}^{\rm
 T}\mathbf{P})^{-1}\mathbf{P}^{\rm T}\tilde{\mathbf{s}},
 \label{binning}
 \end{equation}
where $\mathbf{P}$ is the pointing matrix. It describes the linear
combination coefficients for the (I,Q,U) pixel triplet to produce a
sample of the observed TOD. Each row of the pointing matrix has
three non-zero elements. The vector $\tilde{\mathbf{s}}$ is the
simulated signal TOD (CMB or foreground, see Eq.
(\ref{TOD_estimate2})). The pointing matrix used here will produce a
binned noiseless map that is smoothed with the telescope beam. The
maps were binned from the observations of all four detectors. The
polarisation directions were well sampled in each pixel of the
$N_{\rm side} = 512$ map, leading to 100\% sky coverage and {\it
rcond\/} $\ge$ 0.2165 over the pixels\footnote{The quantity {\it
rcond\/} is the reciprocal of the condition number. {\it rcond\/} is
the ratio of the absolute values of the smallest and the largest
eigenvalue of the $3\times3$ block matrix of a pixel. The matrix
$\mathbf{P}^{\rm T}\mathbf{P}$ is block-diagonal, made up of these
$3\times3$ matrices.}. For the $N_{\rm side} = 1024$ binned
noiseless maps we discarded the pixels with no hits or pixels with
{\it rcond\/} $<$ 0.01. This led to 912,968 unobserved pixels (out
of 12,582,912 pixels of the $N_{\rm side} = 1024$ map).

\begin{figure}[!ht]
    \begin{center}
    \includegraphics[width=\textwidth]{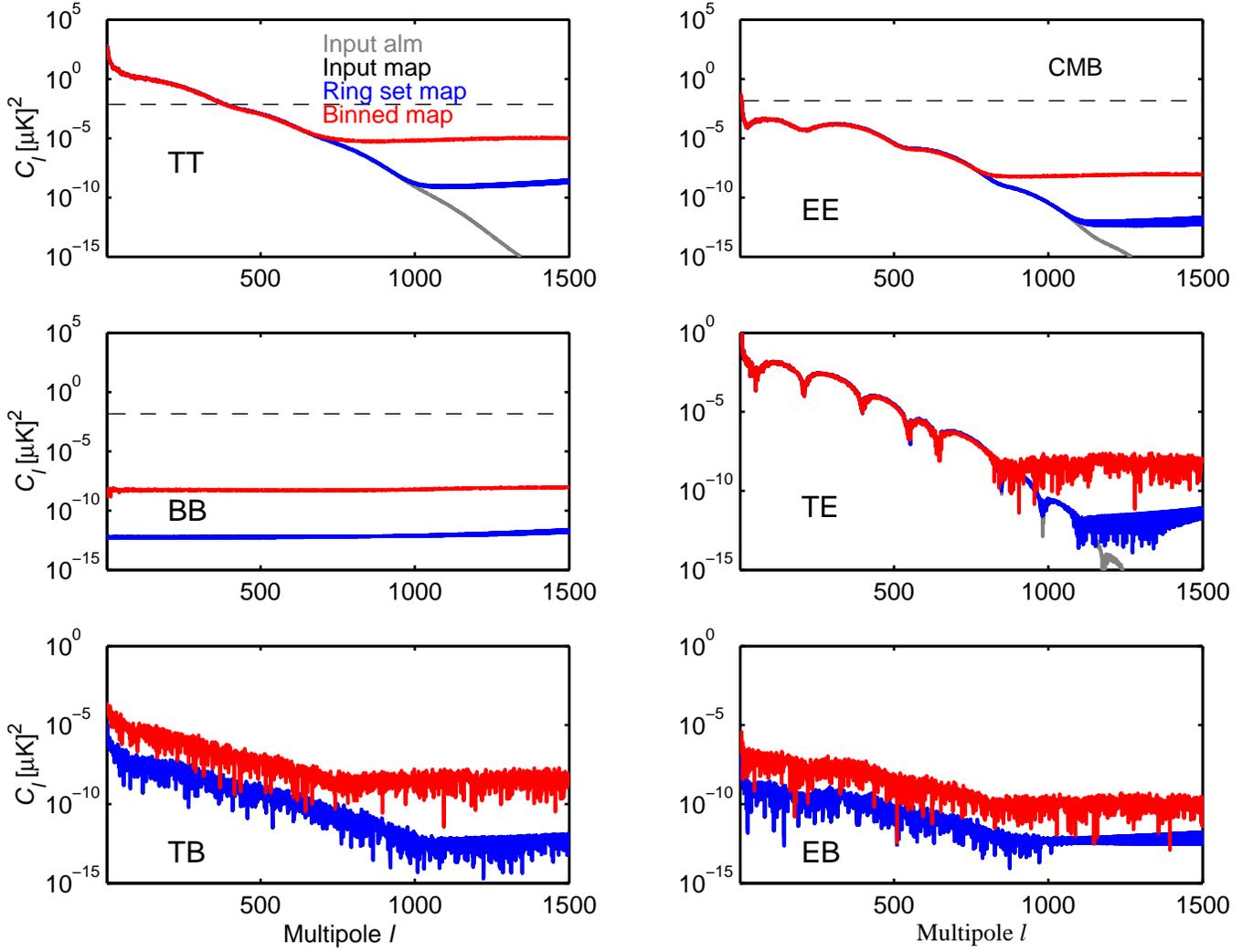}
    \end{center}
  \caption{Angular power spectra of the CMB maps ($N_{\rm side} = 512$):
input map (black curve, not visible since it is so close to the blue
curve), ring set map (blue curve) and the binned noiseless map
(red). The angular power spectrum of the input $a_{\ell m}$ is shown
as well (gray curve). It was calculated in the standard way
$C_\ell^{\rm in} = \sum_{m=-\ell}^{\ell}|a_{\ell m}^{\rm
in}|^2/(2\ell+1)$, where $a_{\ell m}^{\rm in}$ are the input
$a_{\ell m}$.  The horizontal dashed lines give the approximate
spectrum of the white noise map (representing 4 detectors). The
absolute values of the cross correlation spectra are displayed. The
units are CMB microkelvins.}
  \label{fig:cmb_verification}
\end{figure}

\begin{figure}[!ht]
    \begin{center}
    \includegraphics[width=\textwidth]{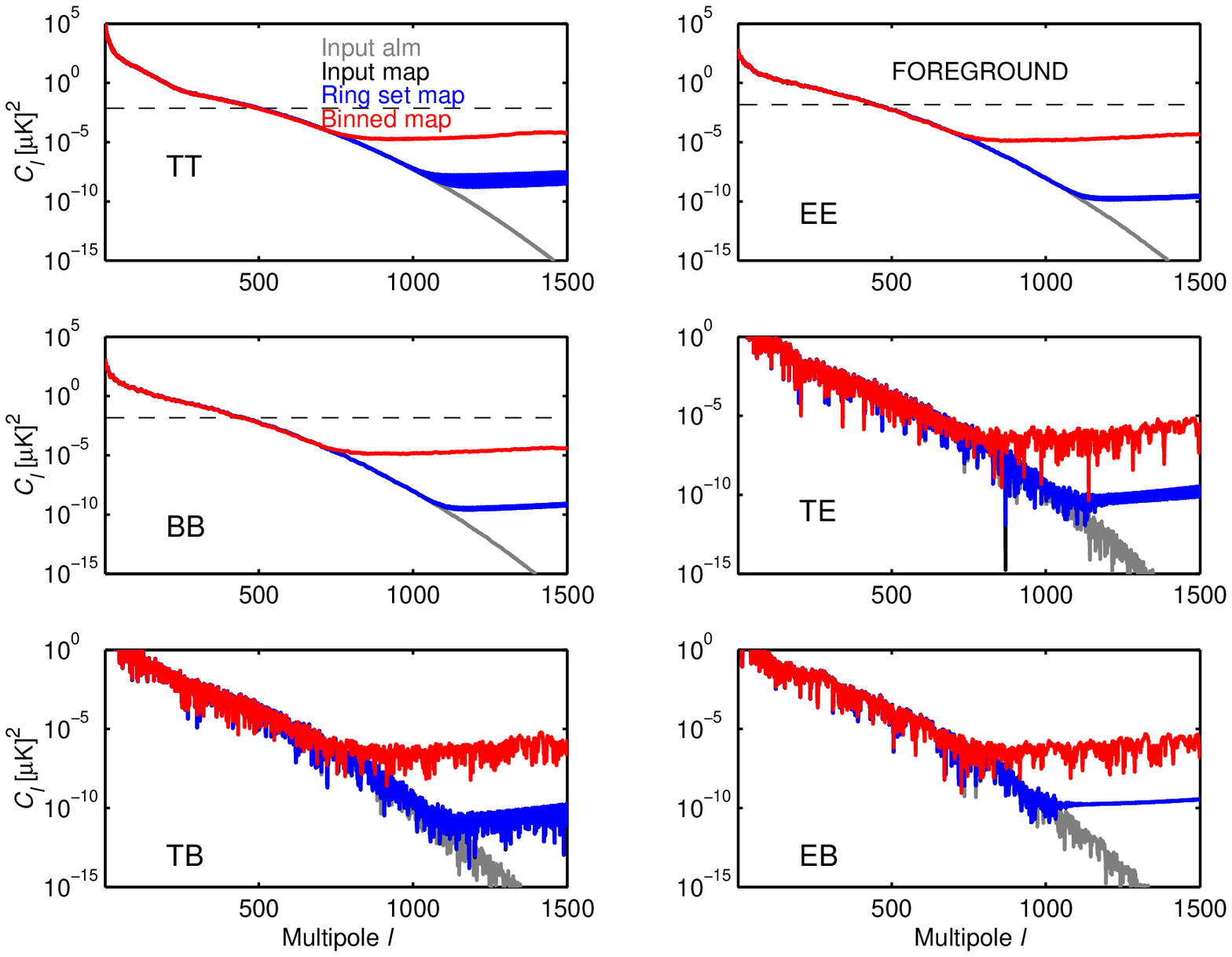}
    \end{center}
  \caption{Same as Fig.~\ref{fig:cmb_verification} but the spectra are for foreground maps.}
  \label{fig:fg_verification}
\end{figure}

The angular power spectra of the CMB and foreground maps are shown
in Figs.~\ref{fig:cmb_verification} and \ref{fig:fg_verification}
(for $N_{\rm side} = 512$). The spectrum of the input $a_{\ell m}$
is shown in gray.  The spectra of the input maps and the ring set
maps are so close to each other that they cannot be distinguished in
the figures (blue curves). The $a_{\ell m}$ coefficients of the
input map were produced using the spherical harmonic transform
(involving numerical integration) over the pixelized celestial
sphere. This discretisation causes a quadrature error in the
numerical integration (G\'orski et al.~\cite{gorski05b}), which
shows up as a high-$\ell$ error floor in the spectrum of the input
map. Therefore the spectrum of the input map deviates from the
spectrum of the input $a_{\ell m}$ at $\ell \gtrsim 1000$.
Decreasing the pixel size will decrease the error floor.  For the
$N_{\rm side} = 1024$ input map the quadrature error floor will
deviate from the spectrum of the input $a_{\ell m}$ at $\ell \gtrsim
1100$. The high-$\ell$ behavior of the spectrum of the ring set map
is also determined by the quadrature error. Therefore the
high-$\ell$ error floor of that spectrum is nearly identical to the
high-$\ell$ error floor of the spectrum of the input map.

The difference between the ring set maps and the input maps is an
indication of the interpolation error that we make in Level-S (see
Sect.~\ref{subsec:convolution}).  The rms of those difference maps
are shown in the first row of Tables~\ref{tab:cmb} and
\ref{tab:foreground}. The difference is very small.

The object of interest in the sky maps is their anisotropy, and the
mean sky temperature is irrelevant.  Therefore, whenever we
calculated a map rms in this study, we subtracted the mean of the
observed pixels from the map before squaring. The rms of a map was
always calculated over the observed pixels only.

\begin{figure}[!ht]
    \begin{center}
    \includegraphics[width=0.6\textwidth]{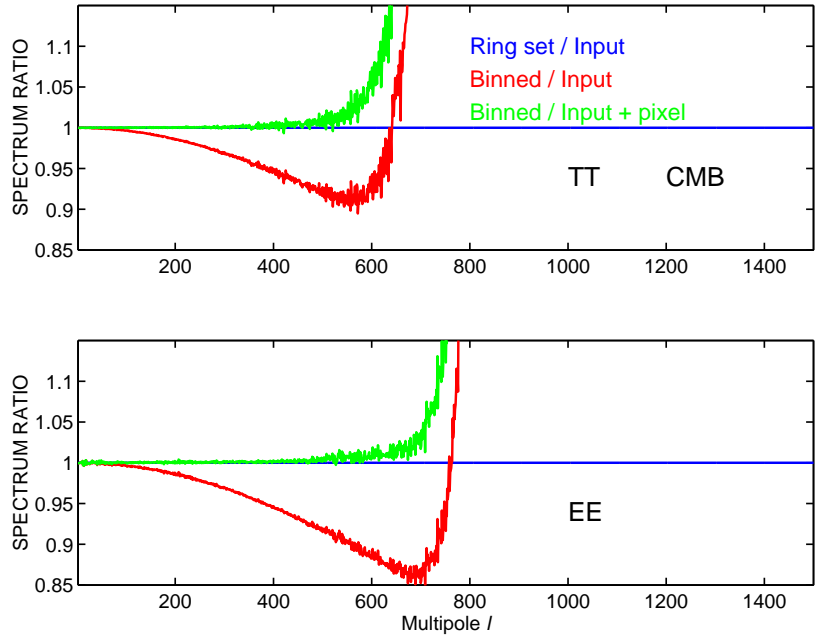}
    \end{center}
  \caption{Ratios of some of the spectra shown in Fig.~\ref{fig:cmb_verification}.
For the green curve the spectrum of the binned noiseless map was
deconvolved with the $N_{\rm side} = 512$ HEALPix pixel window
function before calculating the ratio.
  }
  \label{fig:cmb_ratios}
\end{figure}

\begin{figure}[!ht]
    \begin{center}
    \includegraphics[width=0.6\textwidth]{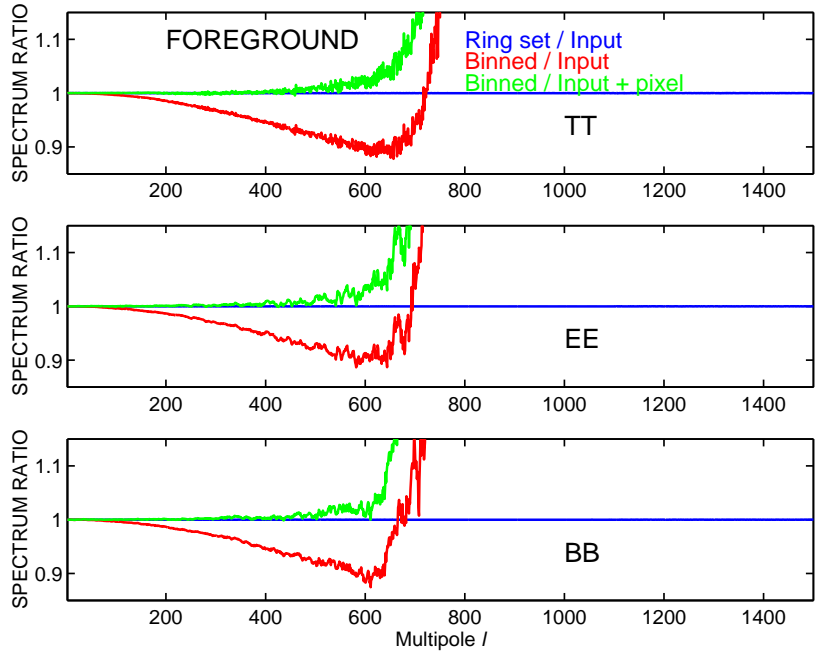}
    \end{center}
  \caption{Ratios of some of the spectra shown in Fig.~\ref{fig:fg_verification}.
For the green curve the spectrum of the binned noiseless map was
deconvolved with the $N_{\rm side} = 512$ HEALPix pixel window
function before calculating the ratio.}
  \label{fig:fg_ratios}
\end{figure}

We also calculated the ratio of the spectra of the $N_{\rm side} =
512$ ring set maps and the input maps. Those ratios are shown in
Figs.~\ref{fig:cmb_ratios} and \ref{fig:fg_ratios} (blue curves).
The spectra are nearly identical. Note, however, that at $\ell
\gtrsim 1000$ the spectra are determined by the quadrature error of
the numerical integration and the ratios do not reflect the
interpolation errors at those $\ell$.  Based on the difference maps
and the spectrum ratios, we conclude that the 11$^{\rm th}$-order
polynomial interpolation that we used in Level-S causes
insignificant error in the simulated TODs.

\begin{table}[h!]
\begin{tabular}{| r  | c c c | c c c |}
\hline
         & \multicolumn{3}{c|}{rms $[\mu$K$]$ ($N_{\rm side} = 512$)} & \multicolumn{3}{c|}{rms $[\mu$K$]$ ($N_{\rm side} = 1024$)}\\
Map difference &   I   &   Q   &   U & I   &   Q   &   U\\
\hline
Ring set - input & 4.070$\times$10$^{-6}$  & 5.786$\times$10$^{-8}$ & 5.149$\times$10$^{-8}$ & 4.070$\times$10$^{-6}$ & 5.771$\times$10$^{-8}$ &5.146$\times$10$^{-8}$ \\
Binned - input &  1.472 & 0.0451 & 0.0458 & 2.291 & 0.0768 & 0.0780\\
Binned - pix convolved input &  1.401 & 0.0415 & 0.0422 & & &\\
\hline
\end{tabular}
\caption{Rms of the CMB difference maps.  All the $N_{\rm side} = 512$
maps and the $N_{\rm side} = 1024$ input and ring set maps cover the full sky,
whereas the $N_{\rm side} = 1024$ binned noiseless map has 912,968 missing
pixels (unobserved or with $rcond< 0.01$). The rms of the
difference maps was determined over the pixels observed by both maps. The
order of the polynomial interpolation is 11 for the ring set maps. For
comparison, the rms of the CMB input map is (I,Q,U) = (84.2, 1.02,
1.03)\,$\mu$K. (The rms of the input map after the convolution with the
$N_{\rm side} = 512$ HEALPix pixel window function is (I,Q,U) = (84.0, 1.00,
1.02)\,$\mu$K.). The units are CMB microkelvins.}
\label{tab:cmb}
\end{table}

\begin{table}[h!]
\begin{tabular}{| r  | c c c |}
\hline
         & \multicolumn{3}{c|}{rms $[\mu$K$]$ ($N_{\rm side} = 512$)} \\
Map difference &   I   &   Q   &   U   \\
\hline
Ring set - input & 3.639$\times$10$^{-5}$ & 7.286$\times$10$^{-6}$ & 7.239$\times$10$^{-6}$  \\
Binned - input & 2.933  & 2.5122 & 2.570  \\
\hline
\end{tabular}
\caption{Same as Table~\ref{tab:cmb}, but this is for the $N_{\rm
side} = 512$ foreground maps. The rms of the foreground input
map is (I,Q,U) = (599.8, 135.3, 133.8) $\mu$K.}
\label{tab:foreground}
\end{table}

The pixel (I,Q,U) triplet of the binned noiseless map represents
(approximately) the mean of the observations falling in that pixel,
whereas the pixel (I,Q,U) of the input map represents an observation
from the pixel center. The non-uniform scatter of observations in
the output map pixels makes the spectrum of the binned noiseless map
different from the spectrum of the input map in two distinct ways.
The scatter of observations causes a spectral smoothing and leads to
an $\ell$ mode coupling (Poutanen et al.~\cite{poutanen06}).

The angular spectra of the binned noiseless maps are shown in
Figs.~\ref{fig:cmb_verification} and \ref{fig:fg_verification} (red curves, for
$N_{\rm side} = 512$). The spectral smoothing is not visible there, but the
high-$\ell$ flat plateau caused by the $\ell$ mode coupling can be seen
clearly.  The plateau is produced when the power from low $\ell$ (where there
is lot of power) is coupled to high $\ell$ (low power) as a result of the
$\ell$-mode coupling.  The {\sc WMAP} team discusses this issue in their 3-year
data analysis and uses the term {\it aliasing\/} for $\ell$-mode coupling
(Jarosik et al.~\cite{jarosik06}). In this study, the flat spectrum plateau of
the $\ell$ mode coupling is called the {\it aliasing error\/}.  Because the two
detectors of a horn have identical circularly symmetric beams and identical
pointings, the high-$\ell$ plateau of the EE and BB spectra is not caused by
the total intensity leaking to polarisation, rather the EE and BB plateaus are
the result of the $\ell$-mode couplings of the E and B mode polarisations.

The horizontal dashed lines of Figs.~\ref{fig:cmb_verification} and
\ref{fig:fg_verification} indicate the approximate level of the white noise
(arising from the four detectors). At $N_{\rm side} = 512$ the aliasing error
(plateau) is clearly smaller than the instrument noise. If the pixel size is
increased, the difference between the noise spectrum and the aliasing error
will decrease, because the increased pixel size will increase the aliasing
error, but the level of the noise spectrum remains nearly unaffected. We made a
binned noiseless foreground map for $N_{\rm side}$ = 64. Its aliasing error had
nearly the same value as the spectrum of the
$N_{\rm side}$ = 64 white noise map.

The spectral smoothing due to the scatter of observations is demonstrated in
Figs.~\ref{fig:cmb_ratios} and \ref{fig:fg_ratios}. The ratios of the spectra
of the binned noiseless map and the input map are shown (red curves, for
$N_{\rm side} = 512$ maps).  For $\ell \lesssim 600$, the spectral smoothing can
be well modelled with the HEALPix pixel window function (G\'orski et
al.~\cite{gorski05b}).  We recalculated the ratios after the spectrum of the
binned noiseless map had been deconvolved with the $N_{\rm side} = 512$ HEALPix
pixel window function.  The resulting ratios are shown as well (green curves).
They show, that, in this case, the spectra (of input map, ring set map and
pixel window deconvolved binned noiseless map) are nearly identical at low
$\ell$ ($\ell \lesssim 400$). The blow-up of the ratios at $\ell \approx 700$
is due to the aliasing error.

The rms of the difference between the binned noiseless maps and the
input maps is shown in Tables~\ref{tab:cmb} and
\ref{tab:foreground}. It contains the effects of both spectral
smoothing and mode coupling. To remove (approximately) the effect of
spectral smoothing we smoothed the input map with the HEALPIx pixel
window function, and then recalculated the difference.  The
resulting rms is shown in the third row of Table~\ref{tab:cmb} (this
was done for the $N_{\rm side} = 512$ CMB input map only). Comparing
the second and the third rows of Table~\ref{tab:cmb}, we see that
the aliasing error (the high-$\ell$ plateau) is the main contributor
in the difference between the binned noiseless map and the input
map.

The main purpose of this paper is to compare map-making algorithms.
We want to isolate errors introduced in the map-making process from
the imperfections (e.g., the scatter of the observations) of the
experimental setup. The difference between the output map and the
input map includes the experimental effects, whereas the difference
between the output map and the binned noiseless map is essentially
free of them.  Therefore we will examine the latter difference maps
in the remaining parts of this paper.

\section{Results}

Simulated TODs for CMB, dipole, foregrounds, and noise were made
individually for the four LFI 30\,GHz detectors.  Maps and angular
power spectra made from combinations of these TODs are discussed
below.  Most of our output maps had $\sim$7 arcmin ($N_{\rm side} =
512$) pixel size.  At this resolution the polarisation directions
were well sampled in every pixel of the map (see
Sect.~\ref{sec:verification}).

MapCUMBA, MADmap, ROMA, and Madam applied the known PSD of the
instrument noise (see Sect.~\ref{subsec:noise}).  In a real
experiment, of course, the noise properties must be estimated from
the observed data.  Here we wanted to avoid the errors that would
arise if we used estimated noise PSDs instead of the actual one. The
length of the uniform baselines was 1~min in Polar and 1.2~s in
\hbox{Madam}.  For comparison, Madam was also run in some cases with
longer 1~min uniform baselines.  Whenever we discuss Madam in this
paper without specifying the baselines, 1.2~s baselines are assumed.

\begin{figure} [!ht]
    \begin{center}
    \includegraphics[scale=0.33,angle=90]{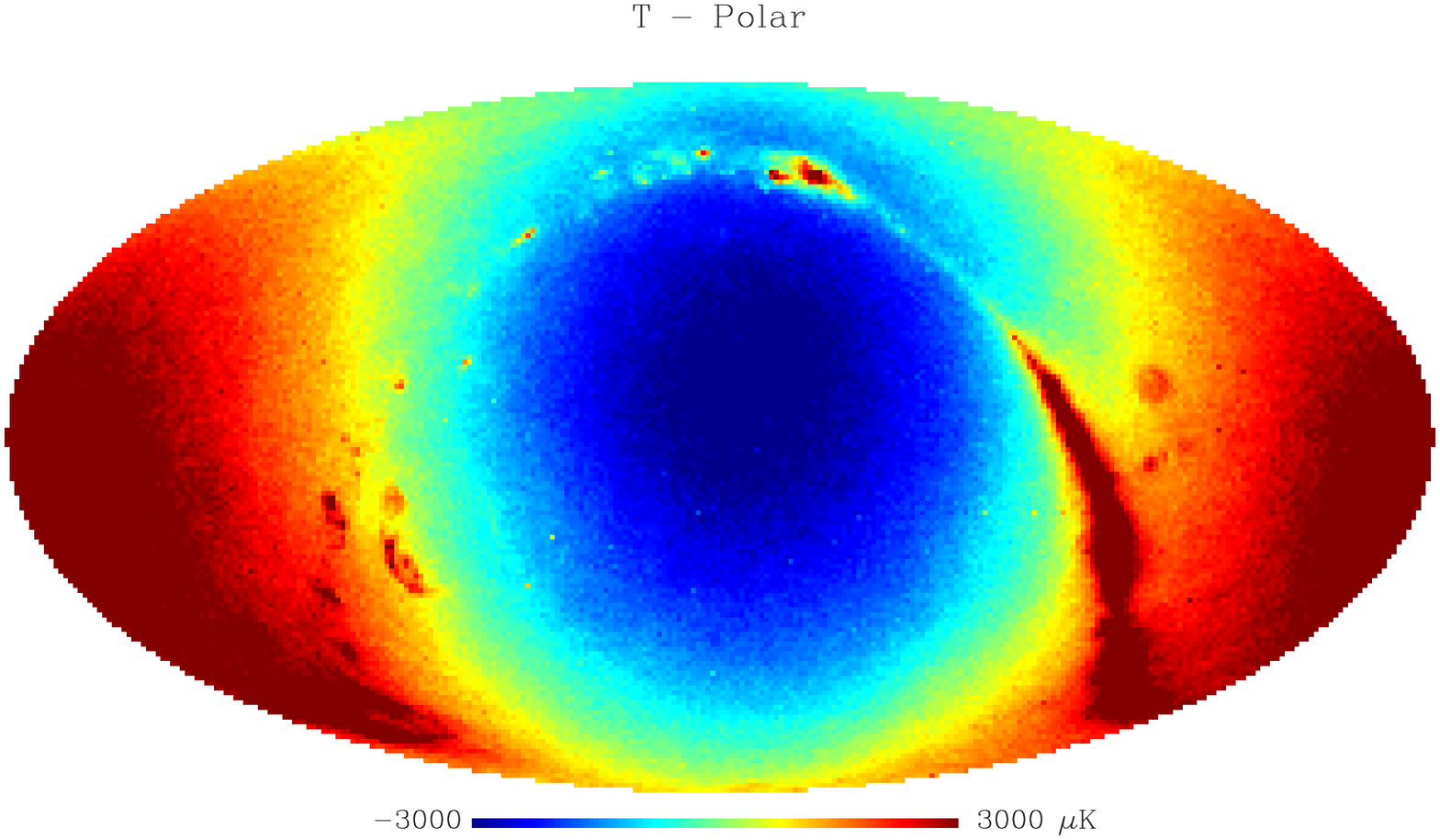}
    \includegraphics[scale=0.33,angle=90]{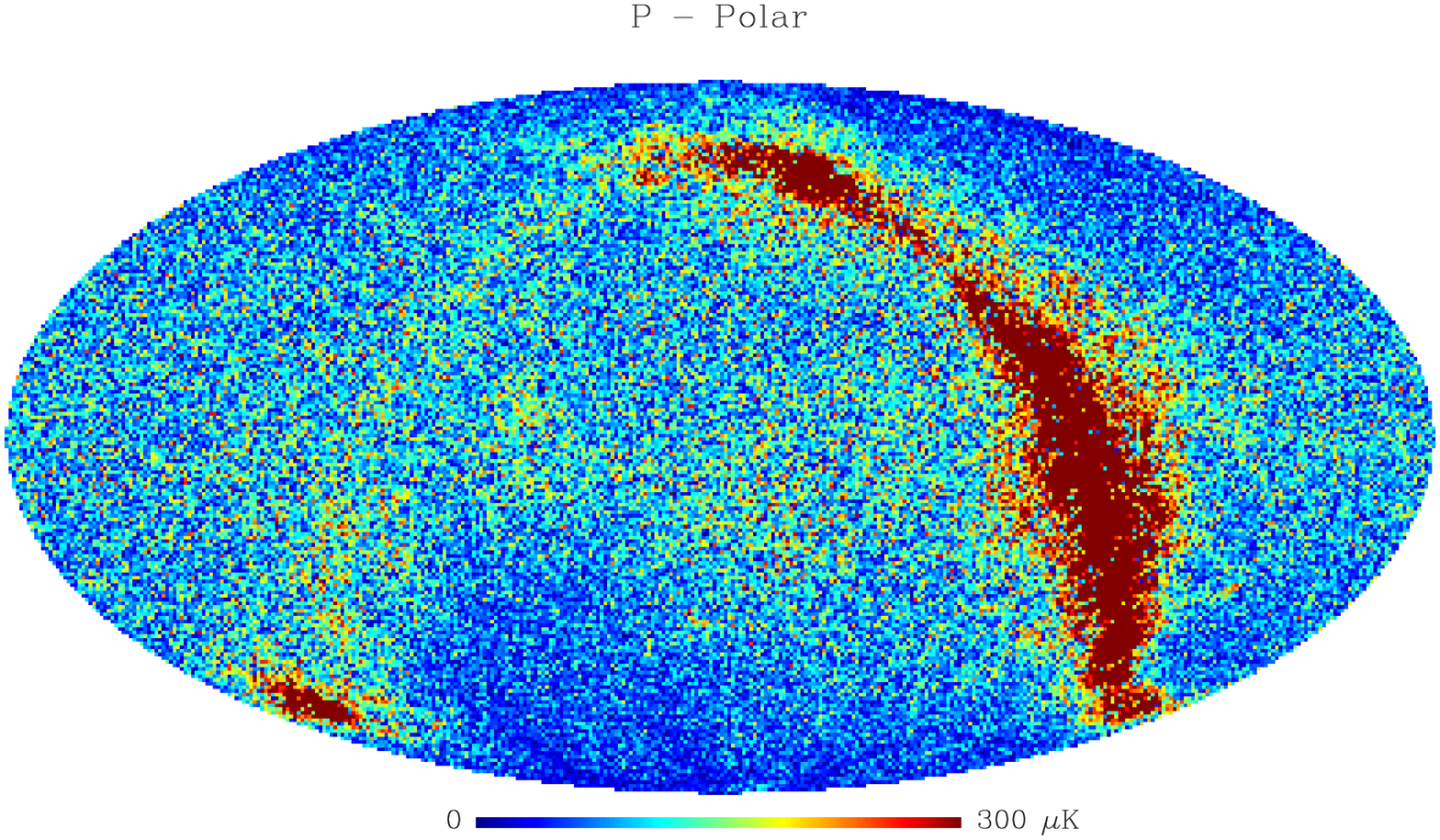}
    \includegraphics[scale=0.33,angle=90]{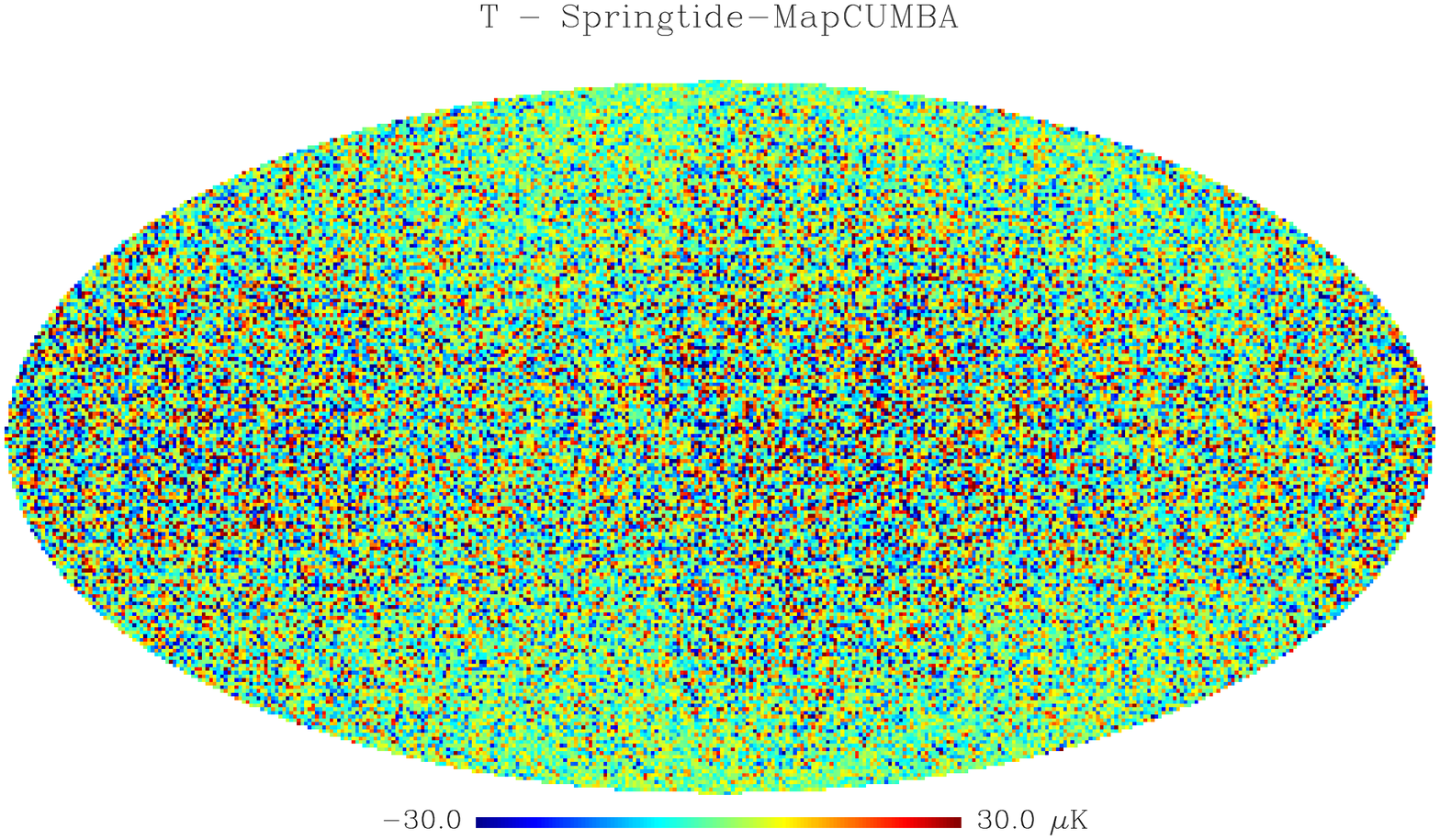}
    \includegraphics[scale=0.33,angle=90]{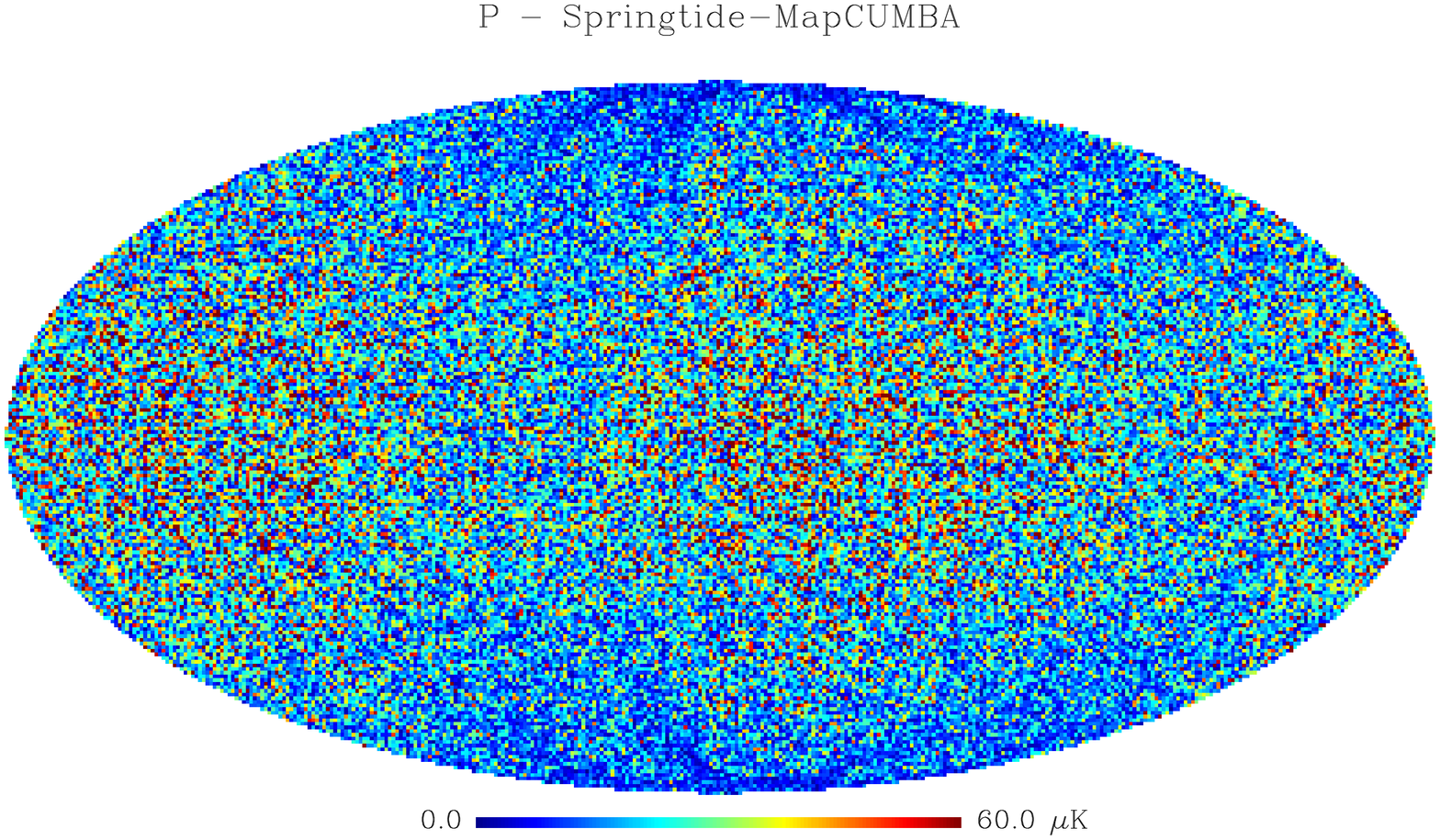}
    \includegraphics[scale=0.33,angle=90]{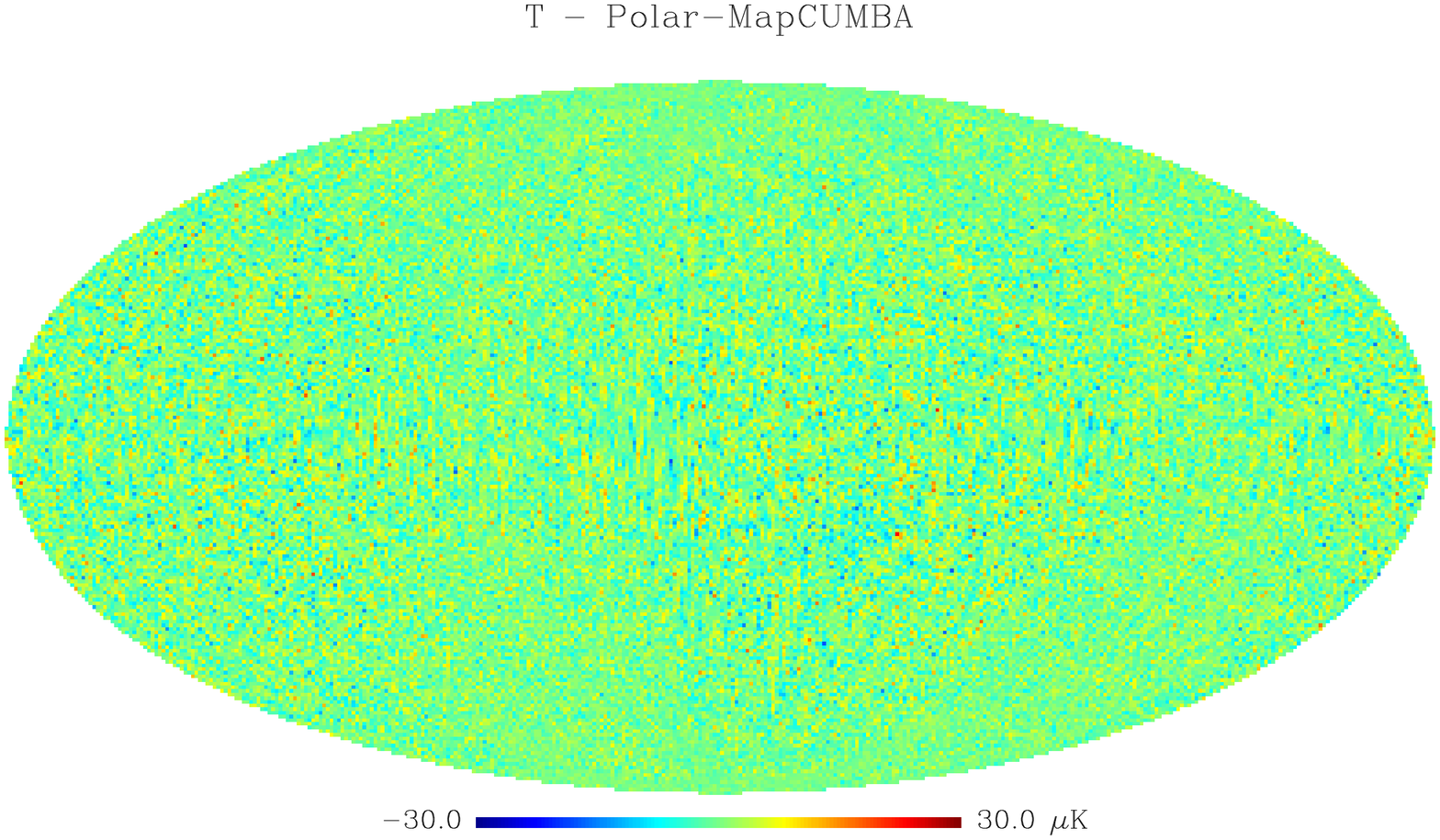}
    \includegraphics[scale=0.33,angle=90]{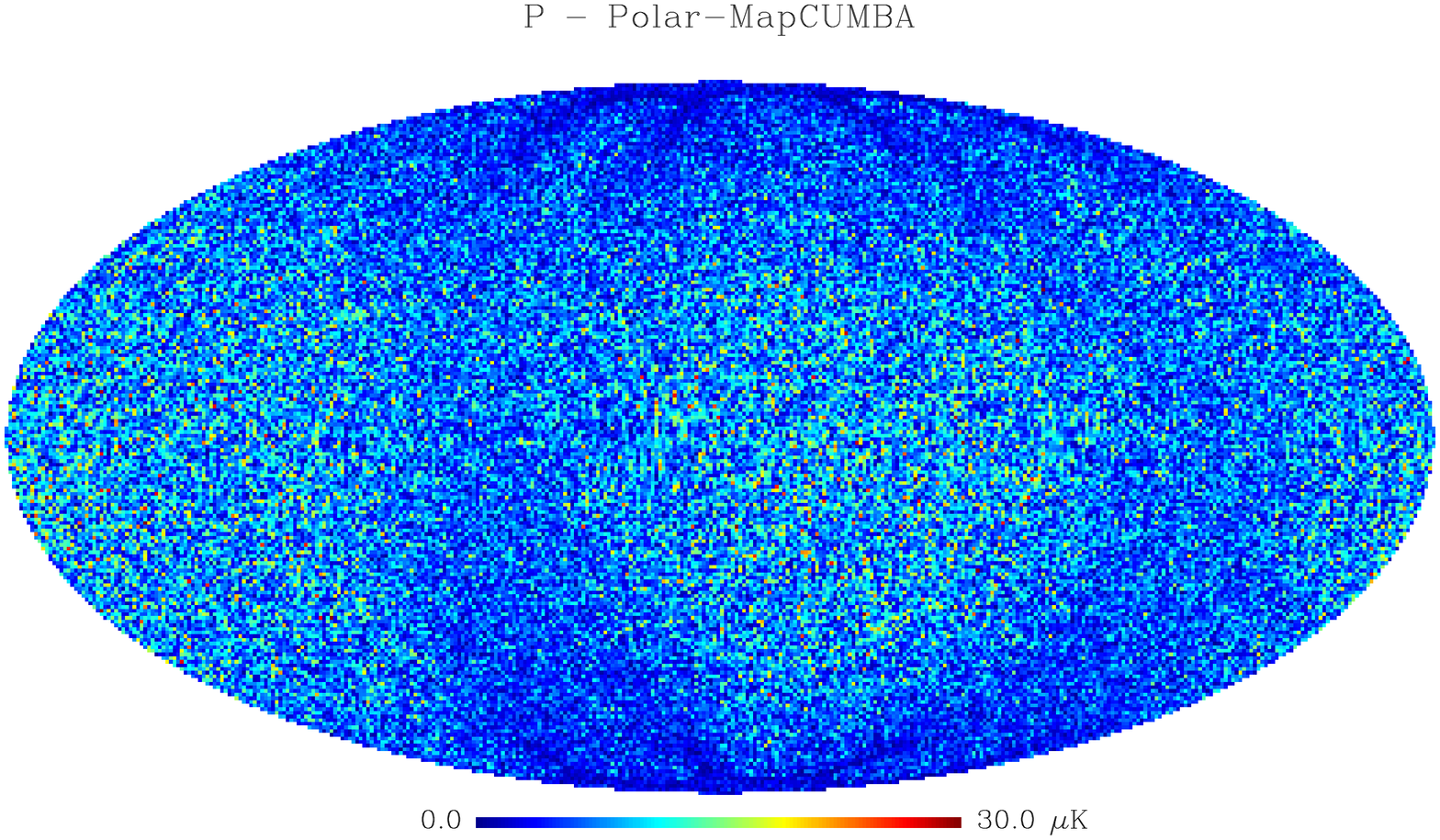}
    \end{center}
\caption{Top row: CDFN output map of Polar. Middle row: CDFN
difference map of Springtide and MapCUMBA. Bottom row: CDFN
difference map of Polar and \hbox{MapCUMBA}. These $N_{\rm side} =
512$ maps contain CMB, dipole, foregrounds, and noise (CDFN). They
are displayed in ecliptic coordinates. The left hand maps are Stokes
I, and the right hand maps are for the magnitude $P$ of the
polarisation vector ($P = \sqrt{Q^2 + U^2}$). The map units are CMB
microkelvins. Note the different color scales of the two bottom $P$
maps. (See www.helsinki.fi/\~{}tfo\_cosm/tfo\_planck.html for
better-quality figures.)}
 \label{fig:cdfn}
 \end{figure}

\begin{figure}[!ht]
    \begin{center}
    \includegraphics[scale=0.45]{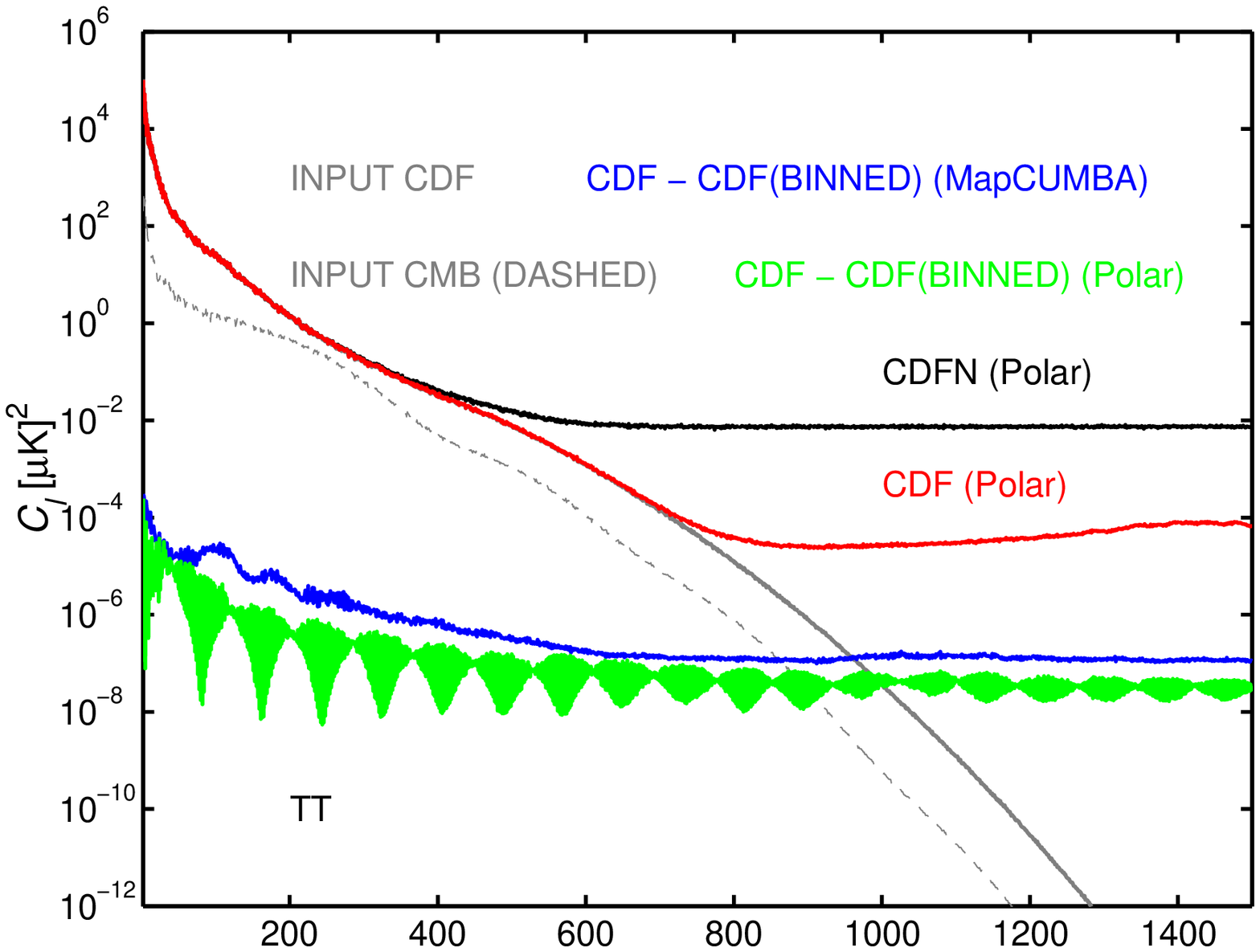}
    \includegraphics[scale=0.45]{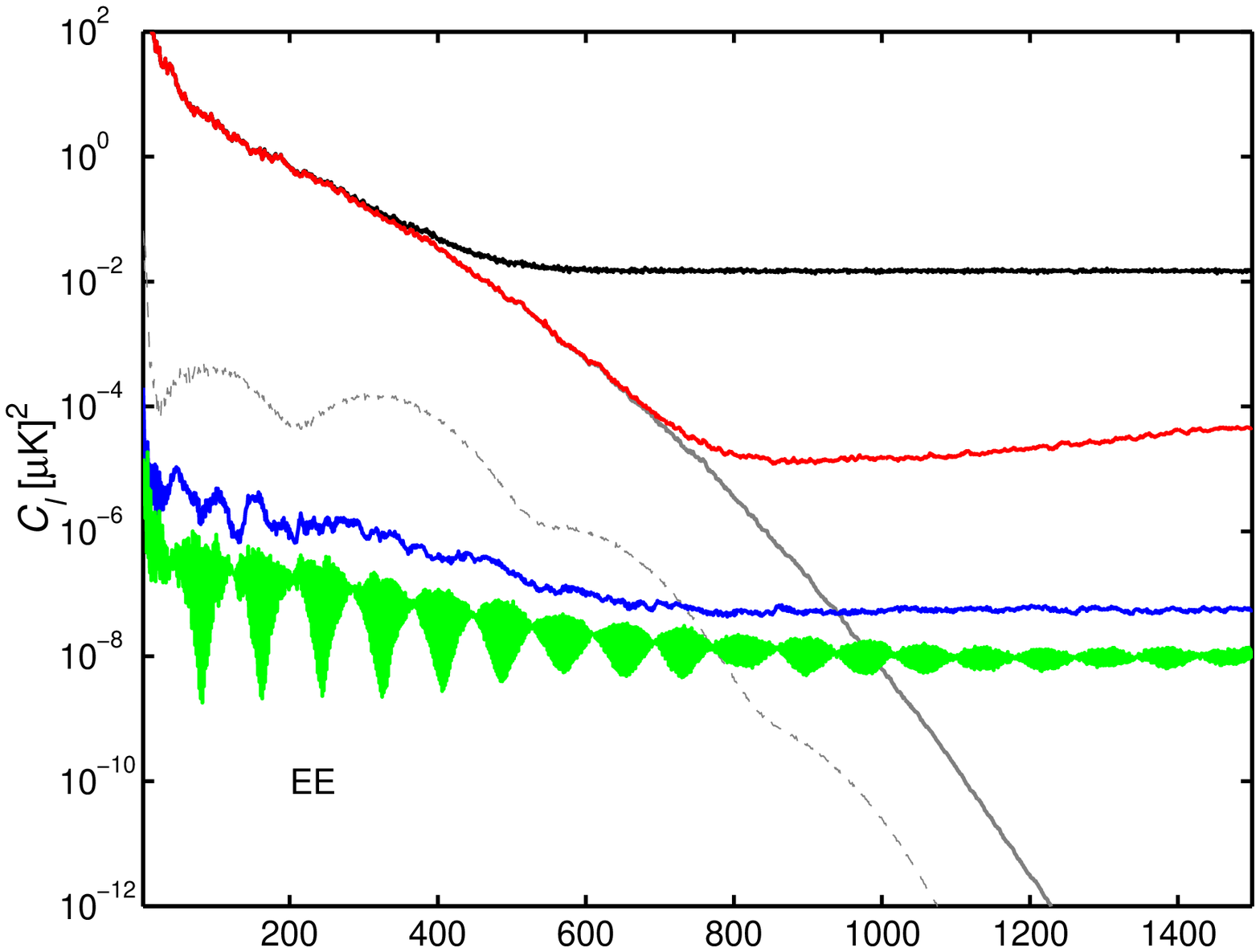}
    \includegraphics[scale=0.45]{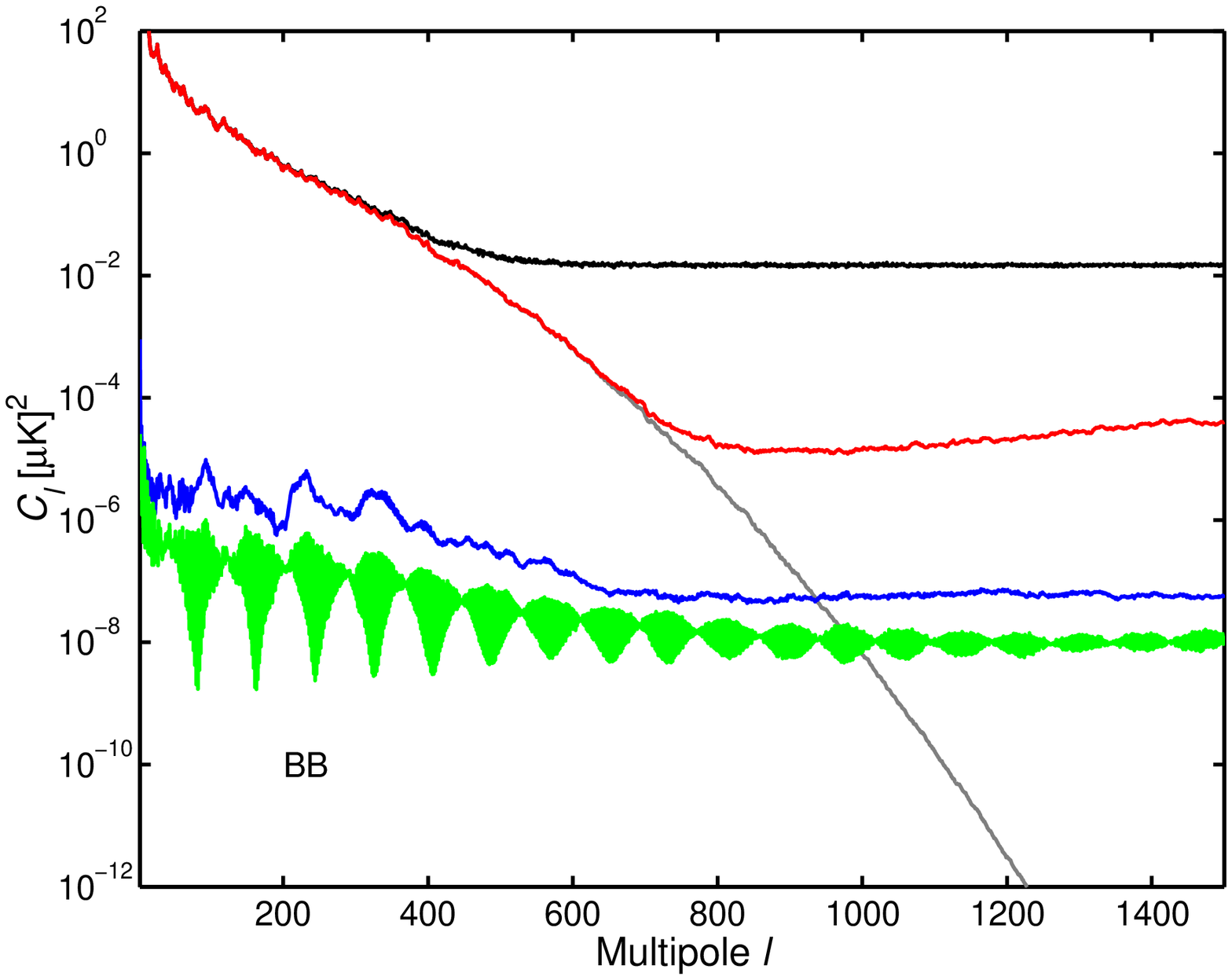}
    \end{center}
\caption{Angular power spectra of the Polar output CDFN
(CMB+dipole+foreground+noise) map (black) and some noiseless (CDF)
maps.  All maps are full sky $N_{\rm side} = 512$ maps . INPUT CDF
(gray) is the spectrum of the sum of the input $a_{\ell m}$ of CMB,
foreground and dipole. INPUT CMB (gray, dashed) is the spectrum of
the input $a_{\ell m}$ of the CMB alone. Note that input $a_{\ell
m}^{\rm B} = 0$ for the \hbox{CMB}. CDF (red) is the same as the
black curve but with no noise. Blue and green curves are the spectra
of the difference maps CDF $-$ CDF(binned), where CDF(binned) is the
binned noiseless map of CDF. Blue curve is for MapCUMBA and green
curve is for Polar. The units are CMB microkelvins.} \label{fig:cdf}
\end{figure}

The top row of Fig.~\ref{fig:cdfn} shows the typical output map
(temperature and polarization) made from the full simulated data
(CDFN = CMB + dipole + foreground + noise), and the black curves in
Fig.~\ref{fig:cdf} show the angular power spectra of this output
map\footnote{Full results are available at
http://spider.ipac.caltech.edu/$\sim$efh/anon$\_$ftp/planck/LFI30.}.
The red curves in Fig.~\ref{fig:cdf} show the corresponding spectra
for the noiseless (CDF) case.  For comparison we show the spectrum
of the input $a_{\ell m}$ too (for CDF and for CMB alone). We see
that the CDFN output map is signal-dominated at large angular scales
($\ell \lesssim 400$) and noise-dominated at small angular scales.
The output map shown is from Polar, but the output maps and their
angular power spectra from all six map-making codes would look the
same in these figures. To bring out the differences we consider
various difference maps and their spectra.

As a first step in characterizing the differences between the
map-making codes, we calculated the pairwise differences between
their output maps. These difference maps are dominated by the
difference in the remaining noise in these output maps. The rms of
these difference maps are given in Table~\ref{tab:cdfndiff}. We see
that the differences between the output maps of the three GLS codes,
as well as Madam, are relatively small.  The Polar and Springtide
maps are more different.  We confirm in
Sect.~\ref{subsec:residual_noise} that the GLS maps contain less
noise than the Polar and Springtide maps.

\begin{table}[h!]
\begin{tabular}{| r | c | c | c | c | c | c |}
\hline
 $N_{\rm side} = 512$, rms $[\mu$K$]$ & Springtide & Polar & Madam & ROMA & MADmap & MapCUMBA\\
\hline
Springtide &   & 14.061 & 14.609 & 14.634 & 14.628 & 14.628\\
Polar &  19.845 &  & 4.042 & 4.116 & 4.104 & 4.104\\
Madam &  20.615 & 5.712 &  & 0.729 & 0.601 & 0.596\\
ROMA &  20.658 & 5.849 & 1.169 &  & 0.409 & 0.415\\
MADmap &  20.642 & 5.801 & 0.833 & 0.822 &  & 0.074\\
MapCUMBA &  20.642 & 5.800 & 0.822 & 0.830 & 0.135 & \\
\hline
\end{tabular}
\caption{Rms of the differences of the CDFN output maps. The data in
the upper right triangle are for the I difference maps. The values
in the lower left triangle are obtained as $\sqrt{(\sigma_{\rm Q}^2
+ \sigma_{\rm U}^2)/2}$, where $\sigma_{\rm Q}$ and $\sigma_{\rm U}$
are the rms of the Q and U difference maps. The maps had $N_{\rm
side} = 512$ and the units are CMB microkelvins.}
\label{tab:cdfndiff}
\end{table}

The Springtide $-$ MapCUMBA and Polar $-$ MapCUMBA difference maps
are shown as the middle and bottom rows of Fig.~\ref{fig:cdfn}. Some
large scale structure (stripes along the scan path) is (barely)
visible in the Polar $-$ MapCUMBA difference map.  These stripes
reflect the difference in the residual $1/f$ noise in the output
maps. Similar large scale structure is not visible in the Springtide
$-$ MapCUMBA difference map because of the higher pixel scale noise
in the Springtide output map.  The rms of the Polar $-$ GLS
difference maps is only $\sim$$1/3.5$ of the rms of the Springtide
$-$ GLS difference maps. The reason for the large Springtide $-$ GLS
output map difference is the high-$\ell$ residual noise of the
Springtide maps as discussed below.

\begin{figure}[!ht]
    \begin{center}
    \includegraphics[scale=0.50]{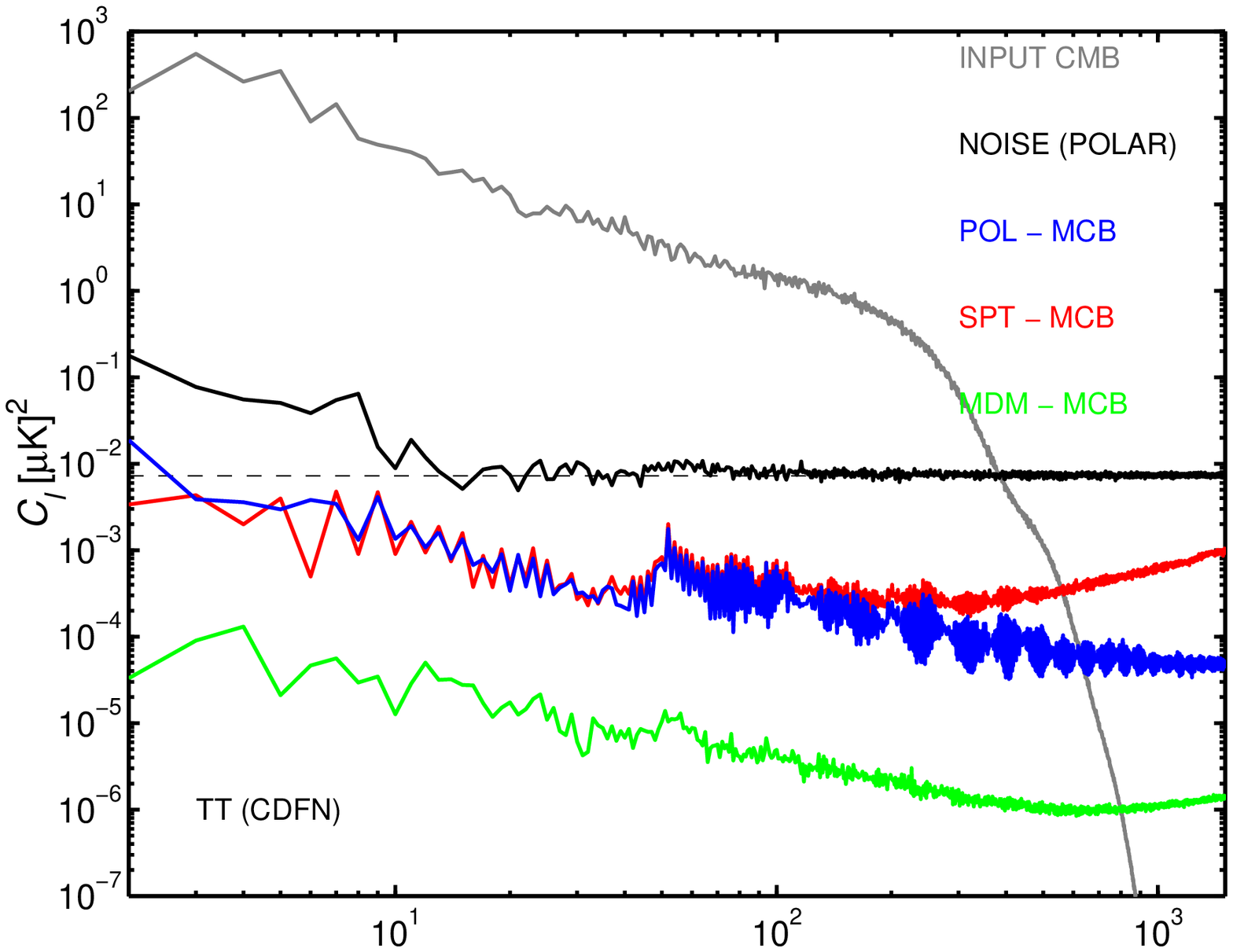}
    \includegraphics[scale=0.50]{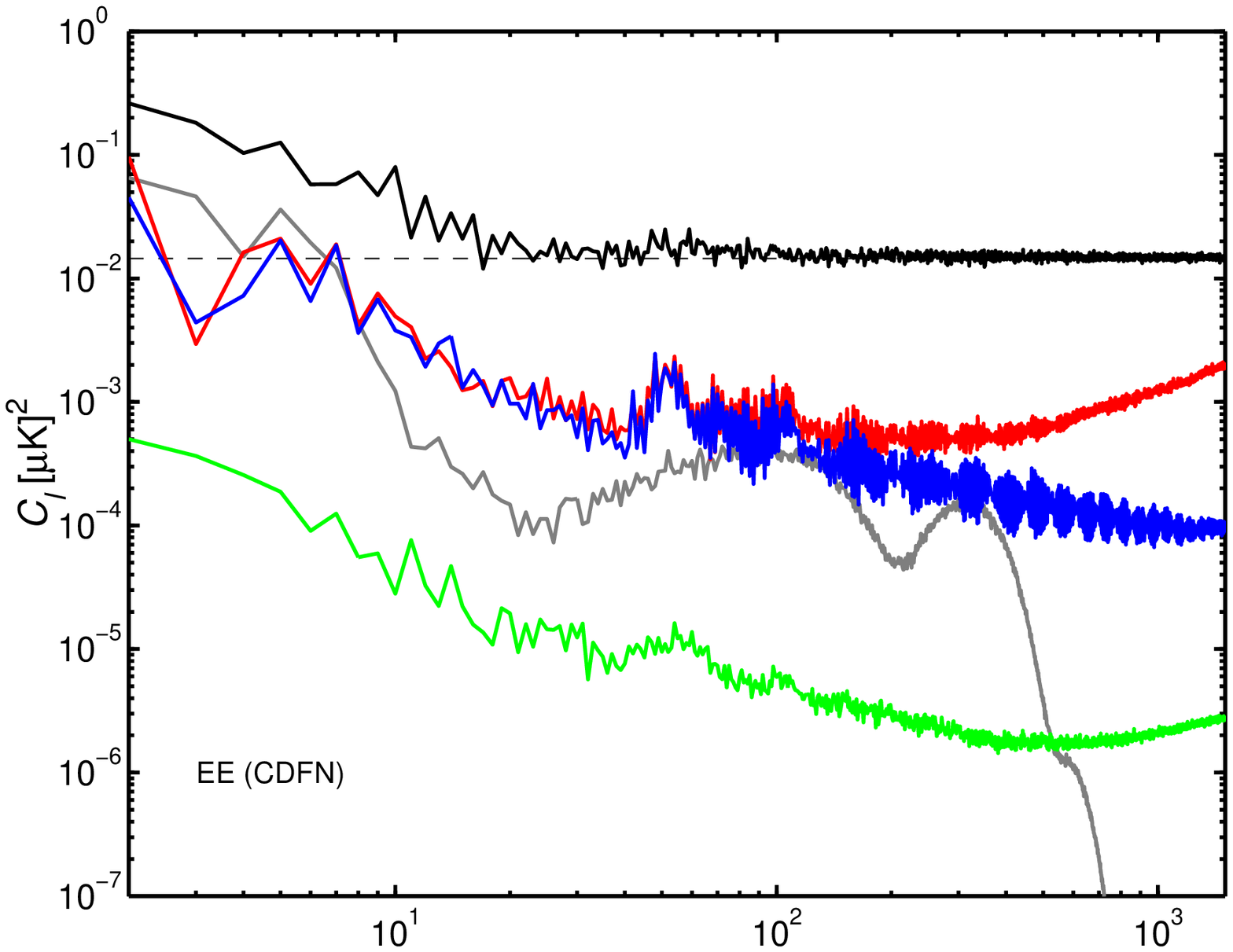}
    \includegraphics[scale=0.50]{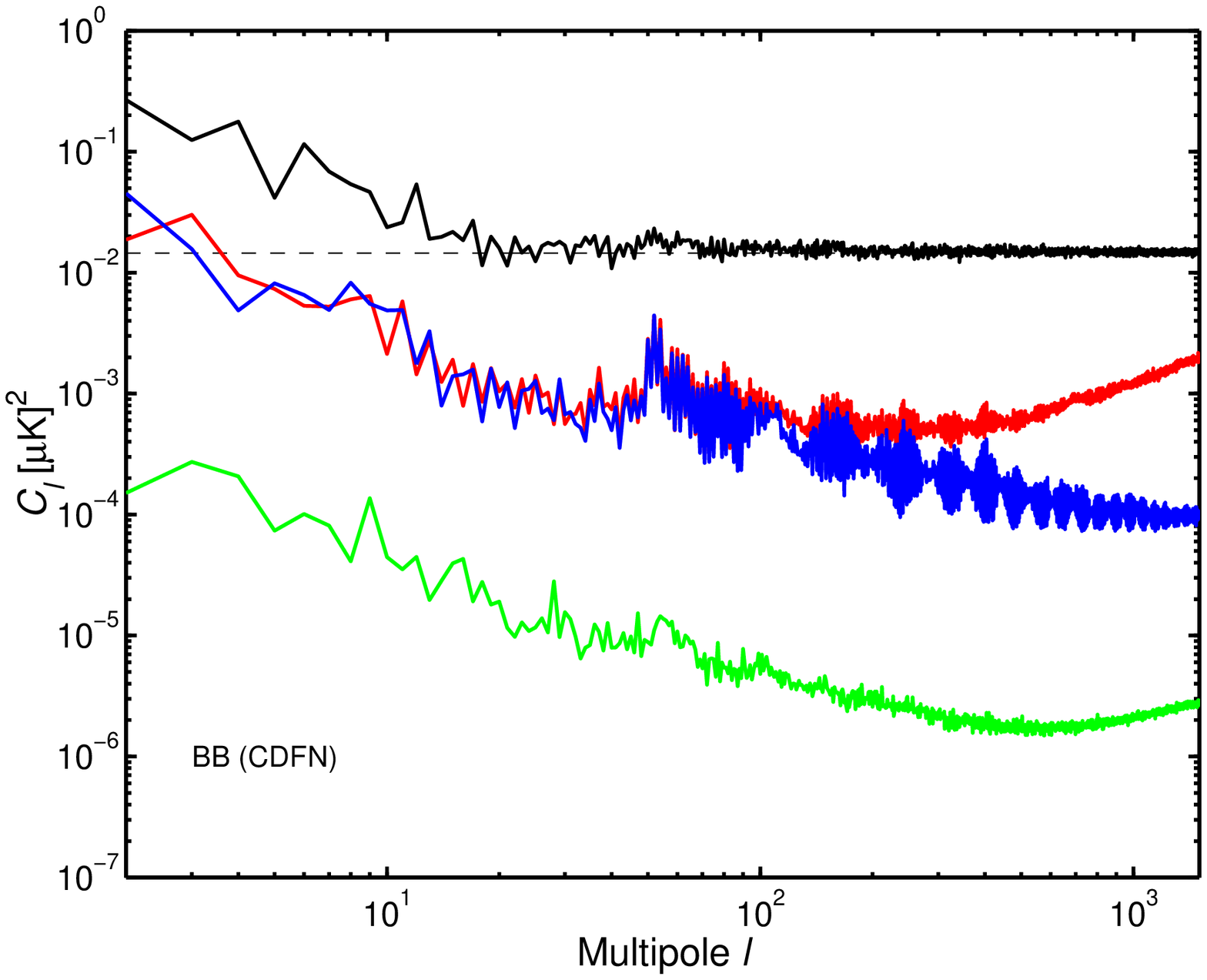}
    \end{center}
\caption{Angular power spectra of the CDFN difference maps. The red
curve is for the difference of the Springtide (SPT) and MapCUMBA
(MCB) output maps, the blue curve is for the difference of Polar
(POL) and MapCUMBA output maps and the green curve is for the
difference of the Madam (MDM) and MapCUMBA output maps. All maps
were $N_{\rm side} = 512$ maps and the units are CMB microkelvins.
For comparison we show the spectra of the CMB input $a_{\ell m}$
(gray curve) and the noise only output map of Polar (black curve).
The horizontal dashed line is an approximation of the spectrum of
the white noise map (see the caption of Table~\ref{tab:n}).}
\label{fig:cdfndiff}
\end{figure}

The angular power spectra of the Springtide $-$ MapCUMBA and Polar
$-$ MapCUMBA CDFN difference maps are shown in
Fig.~\ref{fig:cdfndiff}.  We see from these spectra that the
Springtide and Polar output maps have similar noise structure
(stripes) at large angular scales (low $\ell$), but for Springtide
more noise remains at high $\ell$. This high-$\ell$ noise shows up
as pixel scale noise in the difference map (see
Fig.~\ref{fig:cdfn}). Fig.~\ref{fig:cdfndiff} shows the angular
spectrum of the Madam $-$ MapCUMBA CDFN difference map too. It is
clearly smaller than the other difference map spectra of
Fig.~\ref{fig:cdfndiff}, showing that the noise of the Madam output
map approaches the noise of the GLS output maps.

Fig.~\ref{fig:cdfndiff} further shows, that the CMB temperature
anisotropy signal is larger than the residual noise at low and
intermediate multipoles ($\ell \lesssim 400$, top panel). The
residual noise differences (of different map-making codes) are tiny
fractions of the CMB temperature signal in those multipoles. In the
polarization maps the residual noise differences are of the same
order of magnitude as the CMB signal (middle panel), which is now
significantly smaller than the residual noise and thus difficult to
detect.

The output map from map-making can be represented as a sum of three
components (Poutanen et al.~\cite{poutanen06}): the binned noiseless
map, the residual noise map, and an error map that arises from the
small scale (subpixel) signal structure that couples to the output
map through the map-making. This error map is called the {\it signal
error map\/} in this study. The binned noiseless map is the map we
want and the residual noise map and signal error map are unwanted
errors that depend on the map-making algorithm used.

\subsection{Residual noise maps} \label{subsec:residual_noise}

\begin{figure}[!ht]
    \begin{center}
      \includegraphics[width=0.95\textwidth]{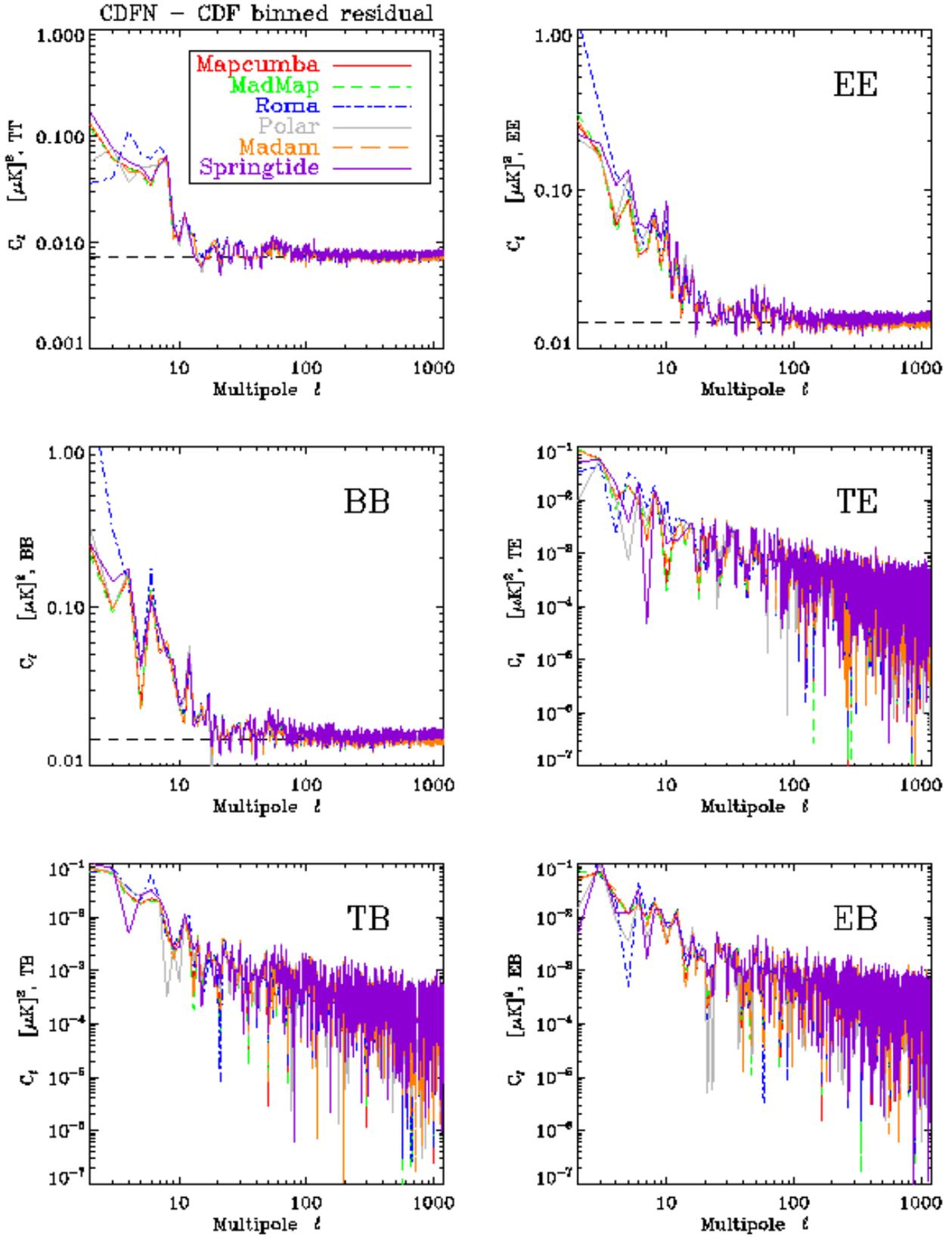}
    \end{center}
\caption{Angular power spectra of the difference maps: CDFN output
map $-$ CDF binned noiseless map. The maps covered the full sky and
their resolution was $N_{\rm side} = 512$. The units are CMB
microkelvins. The horizontal dashed lines in the TT, EE and BB
spectrum plots show the expected angular spectrum of the white noise
map. (See www.helsinki.fi/\~{}tfo\_cosm/tfo\_planck.html for
better-quality figures.)}
  \label{fig:cl_cdfn}
\end{figure}

\begin{figure}[!ht]
    \begin{center}
      \includegraphics[width=0.95\textwidth]{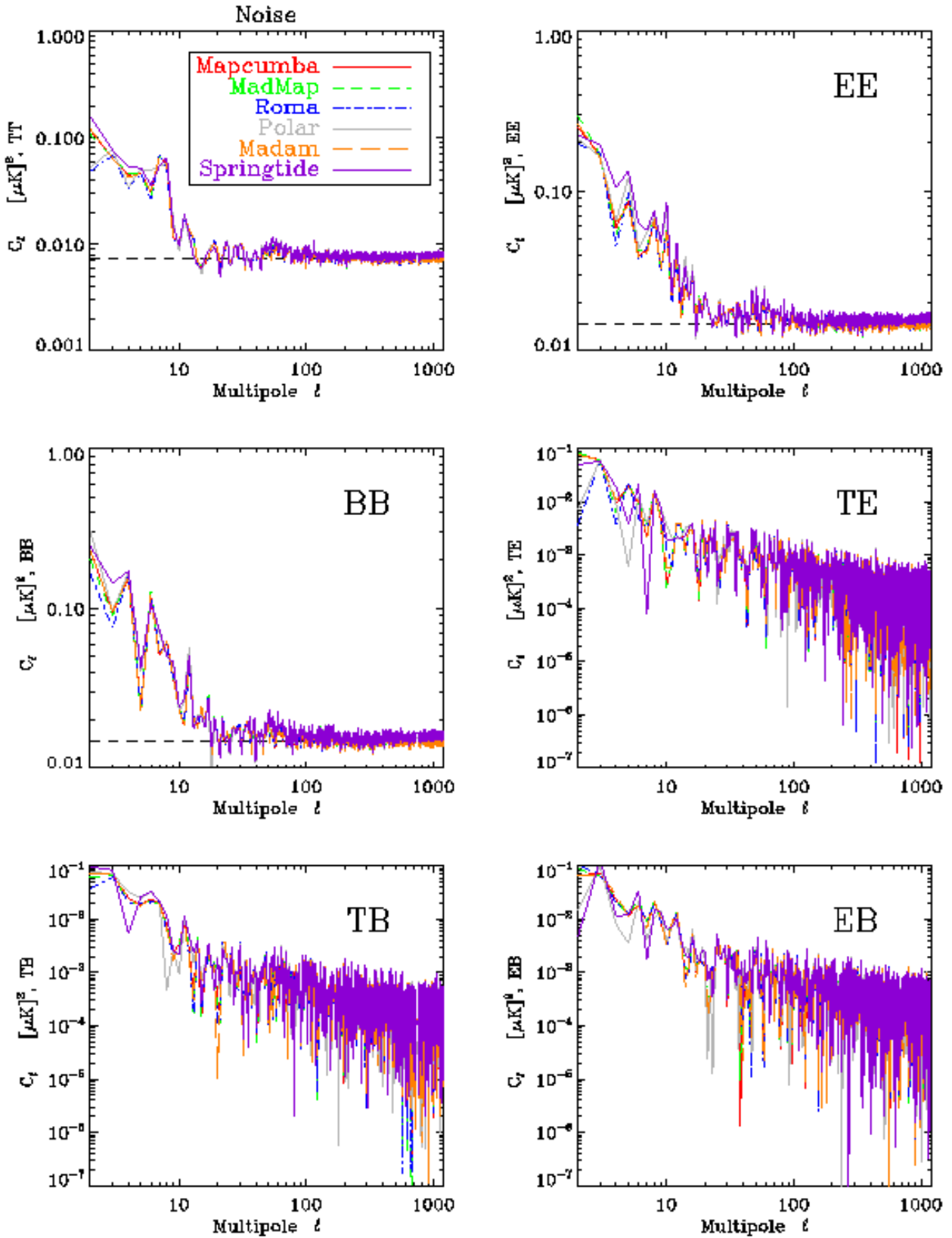}
    \end{center}
\caption{Angular power spectra of the residual noise maps (noise
only output maps). The maps covered the full sky and their
resolution was $N_{\rm side} = 512$. The units are CMB microkelvins.
The horizontal dashed lines in the TT, EE and BB spectrum plots show
the expected angular spectrum of the white noise map. (See
www.helsinki.fi/\~{}tfo\_cosm/tfo\_planck.html for better-quality
figures.)}
  \label{fig:cl_n}
\end{figure}

We computed the total error map (sum of residual noise and signal
error maps) by subtracting the binned noiseless map from the output
map. The angular power spectra of the CDFN total error maps are
shown in Fig.~\ref{fig:cl_cdfn}. These spectra were compared to the
angular power spectra of the residual noise maps, shown in
Fig.~\ref{fig:cl_n}. The residual noise map is made from the noise
TOD only. The residual noise dominates the total error in the CDFN
case.  Therefore the spectra of Figs.~\ref{fig:cl_cdfn} and
\ref{fig:cl_n} are nearly identical. The spectra contain significant
structure at low $\ell$, but tend towards the white noise plateau at
$\ell \gtrsim 20$ (the expected spectra of the white noise map are
shown as dashed horizontal lines). Figs.~\ref{fig:cl_cdfn} and
\ref{fig:cl_n} show small differences between the spectra of
different map-making methods.  The ROMA EE and BB spectra of the
total error are exceptions.  They have somewhat larger power at low
$\ell$ than the other spectra. We could probably have let the ROMA
code to perform some more conjugate gradient iterations for better
recovery of the low-$\ell$ power.

The rms of the residual noise maps (white + $1/f$) and the
approximate level of the $1/f$ noise are given in Table~\ref{tab:n}.
Since map-making algorithms suppress $1/f$ noise, the magnitude of
the residual $1/f$ noise is a good metric for code comparison. The
Polar map contains more residual $1/f$ noise than the maps of the
GLS codes (MapCUMBA, MADmap and ROMA). The difference is, however,
small. The Madam map for 1~min uniform baselines has the same
residual noise as the Polar map. When using short uniform baselines
(1.2~s), Madam can produce maps with as low residual noise as the
GLS codes.

Springtide requires less computing resources than the other codes,
but residual noise from Springtide is higher because it works with
long (1\,hr) ring baselines rather than with the shorter baselines
used in Polar (1~min) and Madam (1.2~s). The higher residual $1/f$
noise of Springtide shows up as a higher map noise (especially at
high $\ell$, as can be seen in Figs.~\ref{fig:cdfn} and
\ref{fig:cdfndiff}).  The larger high-$\ell$ noise of Springtide is
also (barely) visible in the spectra of Figs.~\ref{fig:cl_cdfn} and
\ref{fig:cl_n}.

Table~\ref{tab:n} shows that in the output maps the amount of
residual $1/f$ noise power is 11.6\% (Springtide), 1.9\% (Polar),
and 0.9\% (Madam and GLS codes) of the overall residual noise power.

\begin{table}[h!]
\begin{tabular}{| r  | c c c | c c c |}
\hline
 $N_{\rm side} = 512$ & \multicolumn{3}{c|}{Res. noise, rms $[\mu$K$]$} & \multicolumn{3}{c|} {Res. $1/f$ noise, rms $[\mu$K$]$}\\
Code &   I & Q & U & I & Q & U\\
\hline
Polar &  43.02 & 60.73 & 61.26 & 5.78 & 7.85 & 7.65\\
Springtide &  45.25 & 63.84 & 64.41 & 15.17 & 21.19 & 21.32\\
Madam (1~min) &  43.02 & 60.73 & 61.25 & 5.78 & 7.85 & 7.57\\
Madam (1.2~s) &  42.83 & 60.46 & 60.99 & 4.13 & 5.38 & 5.06\\
MapCUMBA &  42.83 & 60.46 & 60.98 & 4.13 & 5.38 & 4.93\\
MADmap &  42.83 & 60.46 & 60.98 & 4.13 & 5.38 & 4.93\\
ROMA &  42.83 & 60.46 & 60.98 & 4.13 & 5.38 & 4.93\\
\hline
White noise & 42.63 & 60.23 & 60.78 & & &\\
CMB & 84.21 & 1.02 & 1.03 & & &\\
\hline
\end{tabular}
\caption{Pixel statistics of the residual noise maps (output maps of
the noise-only TOD) are shown in the left columns of the table. The residual noise
is a sum of residual white and $1/f$ noise.
For comparison, the rms of the CMB input map and an
approximate rms of a white noise map are also shown. The actual number of hits and
the sampling of the polarisation directions were considered for every pixel when
the white noise rms was computed. The right columns of the table give an
approximate level of the residual $1/f$ noise of the output maps. It is calculated
as $\sqrt{\sigma_{\rm n}^2 - \sigma_{\rm wn}^2}$, where $\sigma_{\rm n}$ is the
rms of the residual noise (from the left columns) and $\sigma_{\rm wn}$ is
the rms of the white noise (from the second bottom row of the table). The units
are CMB microkelvins.}
\label{tab:n}
\end{table}

\subsection{Signal error maps} \label{subsec:signal_error}

If we bin the noiseless signal TOD to a map, scan that map back to
TOD, and subtract it from the original TOD, the resulting difference
TOD gives the {\it pixelisation noise} (Dor\'e et
al.~\cite{dore01}). It was shown in Poutanen et
al.~\cite{poutanen06} that the pixelisation noise is the source of
the signal error map.  Assuming a sum of CMB, dipole, and foreground
signals, four LFI 30\,GHz detectors, 12~months mission time, and
$N_{\rm side} = 512$, we estimated the rms of the pixelisation
noise.  Its value was only about 1\% of the white noise rms
($\sigma$) of a TOD sample (see Sect.~\ref{subsec:noise}). Therefore
the residual noise dominates the total error for $N_{\rm side} =
512$ maps.

In GLS map-making, the pixelisation noise spectrum up to the knee
frequency of the instrument noise contributes to the signal error
map, whereas in destriping only the uniform baselines of the
pixelisation noise contribute (Hivon et al.~\cite{hivon06}; Poutanen
et al.~\cite{poutanen06}). Therefore we expect that the Polar and
Springtide output maps (with their longer baselines) will have
smaller signal error than the GLS output maps. The signal error of
the Madam map (with its short baselines) should approach the signal
error of the GLS maps.

The signal error map can be obtained by subtracting the binned
noiseless map from the signal-only output map. The angular power
spectra of the MapCUMBA and Polar signal error maps (for $N_{\rm
side} = 512$) are shown in Fig.~\ref{fig:cdf}.  The corresponding
spectra of the other GLS codes and Madam would be nearly identical
to the MapCUMBA spectrum, whereas the Springtide spectrum would be
more like the Polar spectrum.  As expected, the GLS signal error is
clearly larger than the signal error of destriping.  The full set of
spectra of the $N_{\rm side} = 512$ signal error maps is shown in
Fig.~\ref{fig:cl_cdf}.

The pixel statistics of the $N_{\rm side} = 512$ signal error maps
for a sum of CMB, dipole, and foreground (CDF) are shown in
Table~\ref{tab:cdf_512}.  Polar and Springtide maps have nearly the
same levels of signal error.  These levels are smaller than the
levels of the GLS and Madam maps. The peak signal error of the Madam
map for 1~min uniform baselines is larger than the peak errors of
the Polar and Springtide maps, although the error rms are nearly the
same.

The levels of the signal error are significantly smaller than the
levels of the residual noise (in $N_{\rm side} = 512$ output maps).
However, the peak signal errors of Table~\ref{tab:cdf_512}
(especially the errors of the GLS and Madam maps) approach or exceed
3\,$\mu$K, which is the LFI performance goal of the maximum
systematic error per pixel (see Table 1.2 in p. 8 of the "{\sc
Planck} Bluebook": Efstathiou et al.~\cite{efstathiou05}).  We need
also to compare the rms of the signal error (left columns of
Table~\ref{tab:cdf_512}) to the rms of the residual $1/f$ noise
(right columns of Table~\ref{tab:n}).  For the GLS codes the signal
error can be up to $\sim$7.4\% of the residual $1/f$ noise. For the
other map-making codes the relative magnitude of the signal error is
smaller.

We examined the signal error maps separately for CMB, dipole, and foreground.  The
resulting statistics are shown in Table~\ref{tab:c+d+f}.  As expected, the
foreground signal (a sum of signals from our own galaxy and extra-galactic point
sources) is the strongest contributor in the signal error map.

\begin{table}[h!]
\begin{tabular}{| r  | c c c | c c c |}
\hline
 $N_{\rm side} = 512$ & \multicolumn{3}{c|}{CDF - CDFbin, rms $[\mu$K$]$} & \multicolumn{3}{c|}{max - min $[\mu$K$]$}\\
Code &   I   &   Q   &   U & I   &   Q   &   U \\
\hline
Polar &  $0.143$ & $0.0746$ & $0.0775$ & $2.46$ & $1.31$ & $1.38$\\
Springtide &  $0.142$ & $0.0744$ & $0.0771$ & $2.39$ & $1.98$ & $1.37$\\
Madam &  $0.290$ & $0.185$ & $0.208$ & $11.90$ & $7.94$ & $9.85$\\
Madam (1~min) &  $0.143$ & $0.0754$ & $0.0783$ & $5.92$ & $3.64$ & $4.16$\\
MapCUMBA &  $0.305$ & $0.193$ & $0.218$ & $14.00$ & $8.50$ & $10.00$\\
MADmap &  $0.304$ & $0.193$ & $0.218$ & $14.00$ & $8.50$ & $10.00$\\
ROMA &  $0.305$ & $0.193$ & $0.218$ & $14.00$ & $8.50$ & $10.00$\\
\hline
\end{tabular}
\caption{Pixel statistics of the noiseless difference maps: output
map $-$ binned noiseless map.  Maps contained CMB, dipole, and
foreground (CDF) and they were full sky maps with $N_{\rm side} = 512$ resolution. The units are CMB microkelvins.} \label{tab:cdf_512}
\end{table}

\begin{table}[h!]
\begin{tabular}{| r  | c c c | c c c | c c c |}
\hline
 $N_{\rm side} = 512$ & \multicolumn{3}{c|}{CMB, rms [nK]} &
\multicolumn{3}{c|}{Dipole, rms [nK]} & \multicolumn{3}{c|}{Foreground,
rms [nK]}\\
Code &   I   &   Q   &   U  &  I   &   Q   &   U  &  I   &   Q   &   U  \\
\hline
Polar &  $75.9$ & $2.53$ & $1.92$ & $18.7$ & $0.170$ & $0.181$ & $119$ & $74.6$ & $77.5$\\
Springtide &  $75.7$ & $2.53$ & $1.92$ &$18.6$ & $0.170$ & $0.181$ & $118$ & $74.3$ & $77.2$\\
Madam &  $143$ & $4.75$ & $3.69$ & $35.5$ & $0.174$ & $0.213$ & $259$ & $185$ & $208$\\
MapCUMBA &  $148$ & $5.11$ & $3.90$ & $36.9$ & $1.97$ & $0.646$ & $273$ & $193$ & $218$\\
MADmap &  $148$ & $4.96$ & $3.77$ & $36.8$ & $0.621$ & $1.36$ & $273$ & $193$ & $218$\\
ROMA &  $148$ & $4.92$ & $3.74$ & $37.3$ & $2.85$ & $0.94$ & $273$ & $193$ & $218$\\
\hline
\end{tabular}
\caption{Pixel statistics of the noiseless difference maps (output
map $-$ binned noiseless map) separately for CMB, dipole, and
foreground.
resolution. Note that the data are now presented in units of CMB
nanokelvins.} \label{tab:c+d+f}
\end{table}

The signal error maps of Polar and Springtide are binned from
uniform baselines that are approximately the baselines of the
pixelisation noise. These baselines represent 1~min circles (Polar)
or 1~hour rings (Springtide) in the sky.  Because the signal errors
of the GLS and Madam maps also contain higher frequency components,
we expect that their signal error maps will contain more small scale
structure than the corresponding maps of Polar and Springtide. The
MapCUMBA and Polar signal error maps for CDF are shown in
Fig.~\ref{fig:cdf_cdfbin}. The largest errors in the MapCUMBA map
are located in the vicinity of the galaxy, where the pixelisation
noise (signal gradients) is strongest.  Such localization of errors
does not exist in the destriped maps, because those codes spread the
errors that arise from the small-scale signal structure over circles
in the sky.

To decrease the signal error, we examined several schemes. They are
described in Sect.~\ref{subsubsec:small_pixels} -
\ref{subsubsec:lowpass_filter}.

\subsubsection{Reducing the pixel size of the output map} \label{subsubsec:small_pixels}

We expect the mean power of the pixelisation noise to decrease as
the pixel size of the output map is decreased. Therefore output maps
with smaller pixels should also have a smaller map-making signal
error. We examined this assumption by making $N_{\rm side} = 1024$
maps from our simulated TODs.

Decreasing the pixel size will also decrease the number of
observations falling in them. In our simulations the $N_{\rm side}
= 1024$ maps had a set of pixels with no observations and another set
with poor sampling of the polarisation directions. The latter pixels
will have ill-behaving $3\times3$ block matrices (see the footnote in
Sect.~\ref{sec:verification}). The residual noise and the signal
error of these pixels will be amplified by the inverses of their
ill-behaving $3\times3$ matrices. The quality of the polarisation sampling
can be quantified by the {\it rcond\/} value of the $3\times3$ matrix (see
the footnote in Sect.~\ref{sec:verification}).  We need to set up a
lower limit for {\it rcond\/} and discard all pixels that have {\it rcond\/} smaller
than this threshold. Our $N_{\rm side} = 1024$ output maps had 191,026 pixels (out of
12,582,912 pixels in total) with no observations.

\begin{figure}[!ht]
    \begin{center}
      \includegraphics[width=0.95\textwidth]{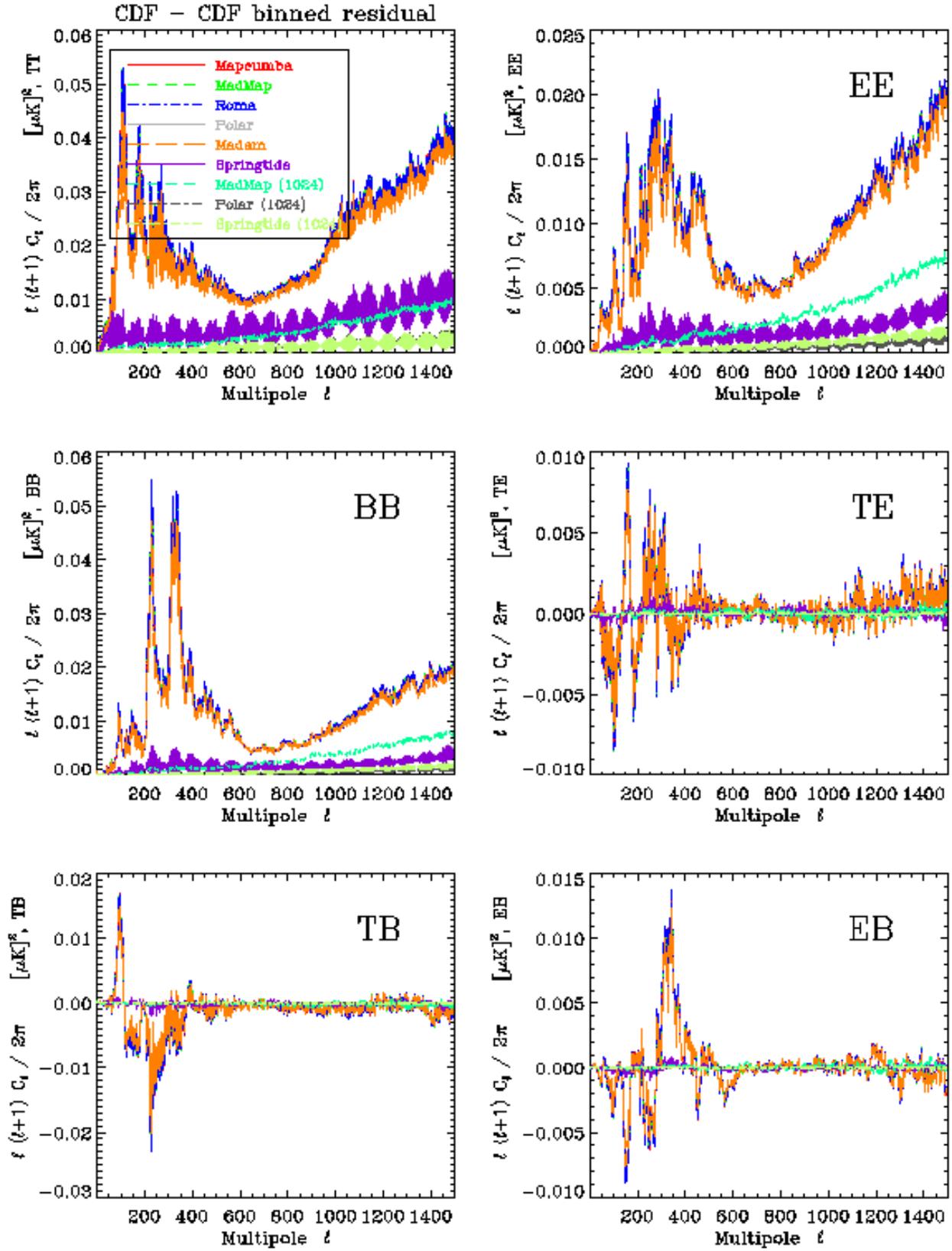}
    \end{center}
\caption{Same as Fig.~\ref{fig:cl_cdfn}, but now for noiseless (CDF)
output maps. All $N_{\rm side} = 512$ maps were full sky maps. Due
to no observations or poor sampling of the polarisation directions
the $N_{\rm side} = 1024$ maps contained a number of unobserved
pixels (see Table~\ref{tab:cdf_1024}). (See
www.helsinki.fi/\~{}tfo\_cosm/tfo\_planck.html for better-quality
figures.)}
  \label{fig:cl_cdf}
\end{figure}

The spectra of the $N_{\rm side} = 1024$ CDF difference maps (binned
noiseless map subtracted from the noiseless output map) are shown in
Fig.~\ref{fig:cl_cdf}. We can see from Fig.~\ref{fig:cl_cdf}, that
decreasing the pixel size (from $N_{\rm side} = 512$ to $N_{\rm
side} = 1024$) decreases the magnitude of the signal error spectrum.
The corresponding map domain statistics for $N_{\rm side} = 1024$
can be found in Table~\ref{tab:cdf_1024}. This table can be compared
to Table~\ref{tab:cdf_512}. This comparison shows, that, although
the rms of most of the $N_{\rm side} = 1024$ signal error maps
are smaller (than the rms of the $N_{\rm side} = 512$ signal error
maps), the peak errors are larger.  Large peak errors were most
likely produced by some ill-sampled pixels that still remain in the
maps.

\begin{table}[h!]
\begin{tabular}{| r  | c c c | c c c |}
\hline
 $N_{\rm side} = 1024$   & \multicolumn{3}{c|}{CDF - CDFbin, rms $[\mu$K$]$} & \multicolumn{3}{c|}{max - min $[\mu$K$]$}\\
Code &   I   &   Q   &   U  & I   &   Q   &   U  \\
\hline
Polar   &  $0.118$ & $0.0722$ & $0.0735$ & $4.63$ & $5.12$ & $5.34$\\
Springtide &  $0.112$ & $0.0820$ & $0.0834$ & $4.11$ & $11.1$ & $11.1$\\
MADmap  &  $0.225$ & $0.185$ & $0.189$ & $16.9$ & $26.3$ & $26.1$ \\
\hline
\end{tabular}
\caption{Pixel statistics of the noiseless difference maps: output
map $-$ binned noiseless map. Maps contained CMB, dipole and
foreground (CDF) and their resolution was $N_{\rm side} = 1024$.
Before computing the statistics we discarded 902,913 pixels from the
Polar and MADmap maps and 837,430 pixels from the Springtide map.
The {\it rcond\/} threshold used in Polar and MADmap was
$\sim$$10^{-2}$, whereas the threshold of the Springtide was
$10^{-3}$. The units are CMB microkelvins.} \label{tab:cdf_1024}
\end{table}

\subsubsection{Discarding crossing points in destriping} \label{subsubsec:discard_crossings}

In destriping, a crossing point is identified, when the samples of
two (or more) baselines measure the same sky pixel. The Polar code
can ignore crossing points that fall outside a sky mask given by the
user.  By using a mask that ignores the galactic and point source
regions, we can prevent the large signal gradients of these areas
from introducing errors in the baseline amplitudes. After computing
the baseline amplitudes, we subtract them from the original TOD and
bin the output map as before. The output map then contains the
observations from the galactic and pointsource regions also.

We used Polar and made $N_{\rm side} = 512$ CDF output maps using
two different masks for discarding the crossing points. We called
the masks {\it Kp0 cut\/} and {\it $\pm$20$\degr$ cut\/}. They are
defined in Fig.~\ref{fig:masks}. The pixel statistics of the signal
error (CDF $-$ CDFbin difference maps) are shown in
Table~\ref{tab:cdf_512_mask}. Kp0 reduces the rms of the signal
error by factors $\sim$1.3 (I map) and $\sim$2.0 (Q and U maps). The
peak errors are reduced by factors $\sim$1.03 (in I map) and
$\sim$2.1 (in Q and U maps).  In spite of the fact that
$\pm$20$\degr$ cut removes larger galactic regions than Kp0, the
signal error of the $\pm$20$\degr$ cut is larger than the signal
error of Kp0. When we introduce a galactic cut in the crossing
points, the magnitude of the pixelisation noise (signal gradients)
becomes smaller, which decreases the signal error. On the other
hand, the galactic cut reduces the number of crossing points (from
the full sky case) and makes the "connection" of our observations
poorer. A poor connection tends to bring more error in the
baselines, which increases the signal error. At some point increase
of the extent of the galactic cut does not help anymore and the
signal error will not decrease.

\begin{table}[h!]
\begin{tabular}{| r  | c c c | c c c |}
\hline
 $N_{\rm side} = 512$ & \multicolumn{3}{c|}{CDF - CDFbin, rms $[\mu$K$]$} & \multicolumn{3}{c|}{max - min $[\mu$K$]$}\\
Polar &   I   &   Q   &   U & I   &   Q   &   U \\
\hline
Kp0 &  $0.108$ & $0.0381$ & $0.0371$ & $2.39$ & $0.68$ & $0.60$\\
$\pm$20$\degr$ &  $0.124$ & $0.0402$ & $0.0400$ & $3.51$ & $0.75$ & $0.63$\\
Full sky &  $0.143$ & $0.0746$ & $0.0775$ & $2.46$ & $1.31$ & $1.38$\\
\hline
\end{tabular}
\caption{Pixel statistics of the noiseless difference maps (signal
error) of Polar: output map - binned noiseless map. Maps contained
CMB, dipole, and foreground (CDF) and they were full sky maps with
$N_{\rm side} = 512$ resolution. The crossing points of baselines
that fall in the galactic or point source regions have been
discarded. For comparison we also show the full sky Polar data (all
crossing points included). It is the same data as in the first row
of Table~\ref{tab:cdf_512}.} \label{tab:cdf_512_mask}
\end{table}

We also made noise-only output maps after masking the crossing
points. The rms of the residual noise map is larger for Kp0 and
$\pm$20$\degr$ cuts than for the noise map with no crossing points
discarded, but the difference is small (see
Table~\ref{tab:n_512_mask}). The goal of destriping is to fit
uniform baselines to the $1/f$ component of the detector noise.
Galactic cut reduces the number of crossing points and makes the
connection of our observations poorer, which leads to a larger
fitting error than when all the crossing points are used. Therefore
the residual noise is larger for the galactic cuts than for the full
sky.

\begin{table}[h!]
\begin{tabular}{| r | c c c |}
\hline
 $N_{\rm side} = 512$ & \multicolumn{3}{c|}{Res. noise, rms $[\mu$K$]$}\\
Polar &   I   &   Q   &   U \\
\hline
Kp0 &  $43.03$ & $60.74$ & $61.27$\\
$\pm$20$\degr$ & $43.04$ & $60.75$ & $61.28$\\
Full sky &  $43.02$ & $60.73$ & $61.26$\\
\hline
\end{tabular}
\caption{Pixel statistics of the Polar residual noise maps. The
crossing points of baselines that fall in the galactic or point
source regions have been discarded. The full sky Polar data (all
crossing points included) is the same data as in the first row of
Table~\ref{tab:n}.} \label{tab:n_512_mask}
\end{table}

\subsubsection{Reducing the pixel size of the crossing points in destriping} \label{subsubsec:pixel_crossings}

In destriping we use a pixelized sky to determine the crossing
points of two (or more) baselines. The samples of the crossing
baselines, that fall in the same pixel, do not necessarily measure
the same point in the sky but they may have different pointings. In
destriping the signal error arises from the differences of these
samples. We expect that reducing the pixel size of the crossing
points will lead to smaller sample differences and thus smaller
signal error. So far in this paper the destriping codes have used
the same pixel size for the crossing points and output maps.

The Polar code has an option, that allows independent pixel sizes
for the crossing points and output map. We made a number of $N_{\rm
side} = 512$ CDF output maps using a different pixel size for the
crossing points. The pixel statistics of the signal error (CDF $-$
CDFbin difference maps) are shown in Table~\ref{tab:cdf_1024_cross}.
It clearly shows, that smaller pixels for the crossing points lead
to smaller signal error.

\begin{table}[h!]
\begin{tabular}{| r  | c c c | c c c |}
\hline
 $N_{\rm side} = 512$ Output map & \multicolumn{3}{c|}{CDF - CDFbin, rms $[\mu$K$]$} & \multicolumn{3}{c|}{max - min $[\mu$K$]$}\\
Polar &   I   &   Q   &   U & I   &   Q   &   U \\
\hline
Crossing point pixels & & & & & &\\
$N_{\rm side} = 4096$ &  $0.0146$ & $0.00737$ & $0.00756$ & $0.31$ & $0.17$ & $0.14$\\
$N_{\rm side} = 2048$ &  $0.0335$ & $0.0187$ & $0.0190$ & $0.93$ & $0.37$ & $0.40$\\
$N_{\rm side} = 1024$ &  $0.0685$ & $0.0364$ & $0.0368$ & $1.59$ & $0.67$ & $0.69$\\
$N_{\rm side} = 512$ &  $0.143$ & $0.0746$ & $0.0775$ & $2.46$ & $1.31$ & $1.38$\\
$N_{\rm side} = 256$ &  $0.393$ & $0.194$ & $0.208$ & $5.67$ & $3.04$ & $3.42$\\
$N_{\rm side} = 128$ &  $1.217$ & $0.570$ & $0.600$ & $17.92$ & $8.50$ & $9.38$\\
$N_{\rm side} = 64$ &  $3.556$ & $1.372$ & $1.448$ & $61.01$ & $18.68$ & $23.92$\\
\hline
\end{tabular}
\caption{Pixel statistics of a number of noiseless difference maps
(signal error) of Polar: output map - binned noiseless map. Maps
contained CMB, dipole, and foreground (CDF) and they were all full
sky maps with $N_{\rm side} = 512$ resolution. We used a different
crossing point pixel size every time we made a map. We discarded
those crossing points that fell in pixels, whose {\it rcond} was $<
10^{-6}$. The data for the $N_{\rm side} = 512$ crossing points is
from the first row of Table~\ref{tab:cdf_512}.}
\label{tab:cdf_1024_cross}
\end{table}

We also made noise-only output maps using a different crossing point
pixel size every time we made a map. The output map resolution was
$N_{\rm side} = 512$ in every map. The pixel statistics of these
residual noise maps are shown in Table~\ref{tab:n_1024_cross} (in
the left I,Q,U columns). We also show the approximate rms of the
residual $1/f$ noise (in the middle I,Q,U columns). They are
determined from the residual noise rms as in Table~\ref{tab:n}.
Smaller crossing point pixels reduce the number of crossing points
and therefore make the connection of our observations poorer. It
leads to a larger error in the baseline fit (cf.
Sect.~\ref{subsubsec:discard_crossings}). Therefore maps with small
crossing point pixels have a higher residual noise than the maps
with larger crossing point pixels (see
Table~\ref{tab:n_1024_cross}).

We approximated the rms of the total effect (residual $1/f$ noise
and signal error) by adding the squares of their rms and taking the
square root of the sum. The result is shown in the right I,Q,U
columns of Table~\ref{tab:n_1024_cross}. We can see, that in this
case the total effect is at its minimum for $N_{\rm side} = 256$ (I)
and $N_{\rm side} = 128$ (Q,U) crossing point pixels. The minima are
not, however, very distinct.

\begin{table}[h!]
\begin{tabular}{| r  | c c c | c c c | c c c |}
\hline
 $N_{\rm side} = 512$ Output map & \multicolumn{3}{c|}{Res. noise, rms $[\mu$K$]$} & \multicolumn{3}{c|}{Res. $1/f$ noise, rms $[\mu$K$]$} & \multicolumn{3}{c|}{Res. $1/f$ noise + signal error, rms $[\mu$K$]$}\\
Polar &   I   &   Q   &   U  &  I   &   Q   &   U  &  I   &   Q   &   U\\
\hline
Crossing point pixels & & & & & & & & & \\
$N_{\rm side} = 4096$ & $43.12$ & $61.00$ & $61.53$ & $6.48$ & $9.66$ & $9.58$ & $6.48$ & $9.66$ & $9.58$\\
$N_{\rm side} = 2048$  & $43.08$ & $60.84$ & $61.37$ & $6.21$ & $8.59$ & $8.49$ & $6.21$ & $8.59$ & $8.49$\\
$N_{\rm side} = 1024$ & $43.04$ & $60.76$ & $61.29$ & $5.93$ & $8.01$ & $7.89$ & $5.93$ & $8.01$ & $7.89$\\
$N_{\rm side} = 512$  & $43.02$ & $60.73$ & $61.26$ & $5.78$ & $7.78$ & $7.65$ & $5.78$ & $7.78$ & $7.65$\\
$N_{\rm side} = 256$ & $43.00$ & $60.70$ & $61.22$ & $5.63$ & $7.54$ & $7.33$ & $5.64$ & $7.54$ & $7.33$\\
$N_{\rm side} = 128$  & $42.99$ & $60.69$ & $61.22$ & $5.55$ & $7.46$ & $7.28$ & $5.68$ & $7.48$ & $7.31$\\
$N_{\rm side} = 64$ & $42.99$ & $60.69$ & $61.21$ & $5.55$ & $7.46$ & $7.24$ & $6.59$ & $7.58$ & $7.39$\\
\hline
\end{tabular}
\caption{Pixel statistics of a number of Polar residual noise maps
(left I,Q,U columns). All maps had $N_{\rm side} = 512$ resolution,
but we used a different crossing point pixel size every time we made
a map. An approximation of the rms of the residual $1/f$ noise is
also shown (middle I,Q,U columns). It was determined as in
Table~\ref{tab:n}. We approximated the rms of the total error
(residual $1/f$ noise and signal error) by adding the squares of
their rms and taking the square root of the sum. The rms to be added
were taken from Table~\ref{tab:cdf_1024_cross} and from the middle
I,Q,U columns of this table. The result is shown in the right I,Q,U
columns.} \label{tab:n_1024_cross}
\end{table}

The effect of the crossing point pixel size on the destriping errors
has been studied elsewhere too. Larqu\`{e}re~(\cite{larquere06})
compares the baseline amplitudes that are determined from the TODs
of CMB, white noise or their sum. This study reveals, that
decreasing the size of the crossing point pixels will decrease the
amplitudes of the baselines determined from the CMB TOD and increase
the baseline amplitudes of the white noise TOD. The opposite will
occur if the crossing point pixel size is increased. The results of
Larqu\`{e}re~(\cite{larquere06}) concur with ours.

The existing GLS map-making codes do not allow us to use different
pixel sizes for the crossing points and output map. Therefore we
could not examine this approach in the GLS codes. In principle one
could use different pixel sizes in the GLS codes too, but that would
require a major rewriting of the existing codes. Appendix A of
Poutanen et al.~\cite{poutanen06} shows a possible way to do this.
It is shown there, that instead of solving the output map from the
usual GLS map-making equation (A1) one can use Eqs. (A6) and (A7)
and obtain the same output map. In this approach one first
determines a TOD domain estimate of the correlated part of the noise
(vector $\Delta$ in Eq. (A7)), then subtracts it from the original
TOD and finally bins the output map from the difference (as done in
Eq. (A6)). One could use smaller pixel size for the noise estimate
than for the output map. Solving the estimate of the correlated
noise from Eq. (A7) in GLS map-making and solving the baseline
amplitudes in destriping are closely related operations. They both
use the differencies of the observations of the same points of the
sky taken at different times.

\subsubsection{Filtering out pixel noise} \label{subsubsec:lowpass_filter}

As discussed in Dor\'e et al.~\cite{dore01} and Poutanen et
al.~\cite{poutanen06}, the pixelisation noise is due to the presence
of signal at scales smaller than the pixel size, while the map-making
modelisation usually assumes the signal to be uniform within each pixel.
It is therefore tempting to treat this pixelisation noise on the same footing as
the instrumental noise by casting the data stream $d$ as
\begin{equation}
  d = \Point m + n_i + n_p
\end{equation}
where $\Point$ is the pointing matrix, $m$ the {\em pixelised} sky
signal, $n$ the instrumental noise and $n_p$ the pixelisation noise
(ie, the difference between a data stream obtained when scanning the
true sky, to one obtained when scanning the pixelised map). As
summarised in Ashdown et al.~\cite{ashdown06}, the optimal map
equation reads
\begin{equation}
  m = (\Point^T \Noise^{-1} \Point)^{-1} \Point^T \Noise^{-1} d
  \label{eq:optmap}
\end{equation}
where the time-time noise correlation matrix $\Noise$ is
\begin{equation}
  \Noise_{tt'} = \Noise_{tt'}^{(i)} + \Noise_{tt'}^{(p)}
  = \left\langle n_i(t) n_i(t') \right\rangle + \left\langle n_p(t) n_p(t') \right\rangle,
\end{equation}
assuming that $n_i$ and $n_p$ are not correlated. This modelisation
of the pixelisation noise ensures that it will be optimally weighted
down during the map-making, without biasing the map obtained.

While the instrumental noise frequency power spectrum for the detector
considered is given by
\begin{equation}
  N^{(i)}(f) = 1 + (f/0.05)^{-1.7},
\end{equation} where $f$ is the
frequency in Hz, we have assumed the power spectrum of the
pixelisation noise to be
\begin{equation}
  N^{(p)}(f) = f/5.
  \label{eq:pn_model}
\end{equation}
 Using this noise filter,
we have constructed the optimal map (Eq.~\ref{eq:optmap}) of the
noiseless data stream. However, this maneuver is not enough to
reduce the pixelisation noise left on the map. The reason is that
while the ansatz chosen in Eq.~\ref{eq:pn_model} describes the
sub-pixel power encountered while the detector scans across the
pixel (at a time scale of a few milliseconds), it does not take into
account the sub-pixel power encountered between different visits of
the same pixel (at time scales ranging from one minute to a few
months for Planck). A more sophisticated treatment of the
pixelisation noise is therefore necessary.

\subsection{Galactic cut of the output map} \label{subsec:galactic_cut}

\begin{figure} [!ht]
    \begin{center}
    \includegraphics[scale=0.33,angle=90]{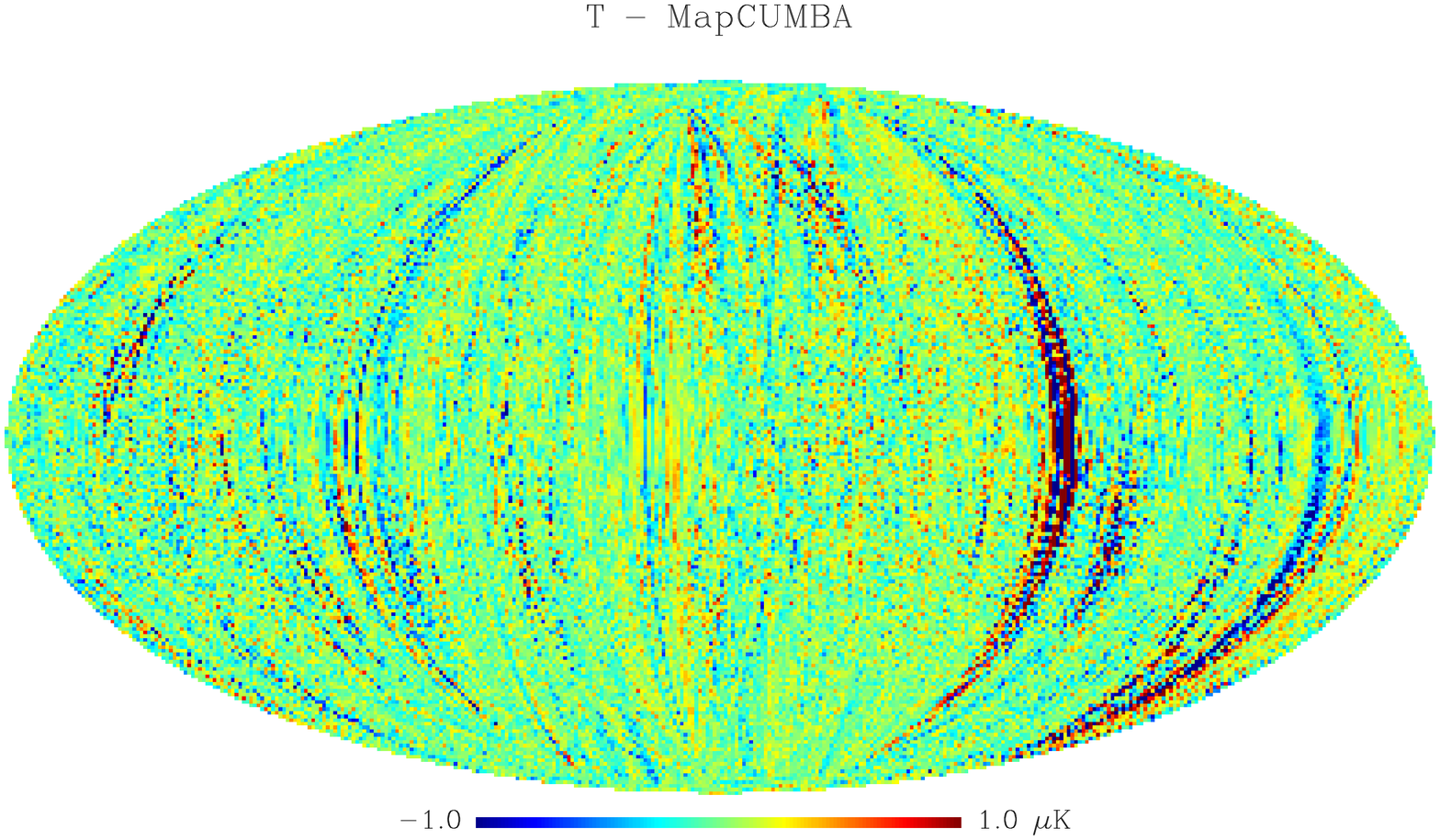}
    \includegraphics[scale=0.33,angle=90]{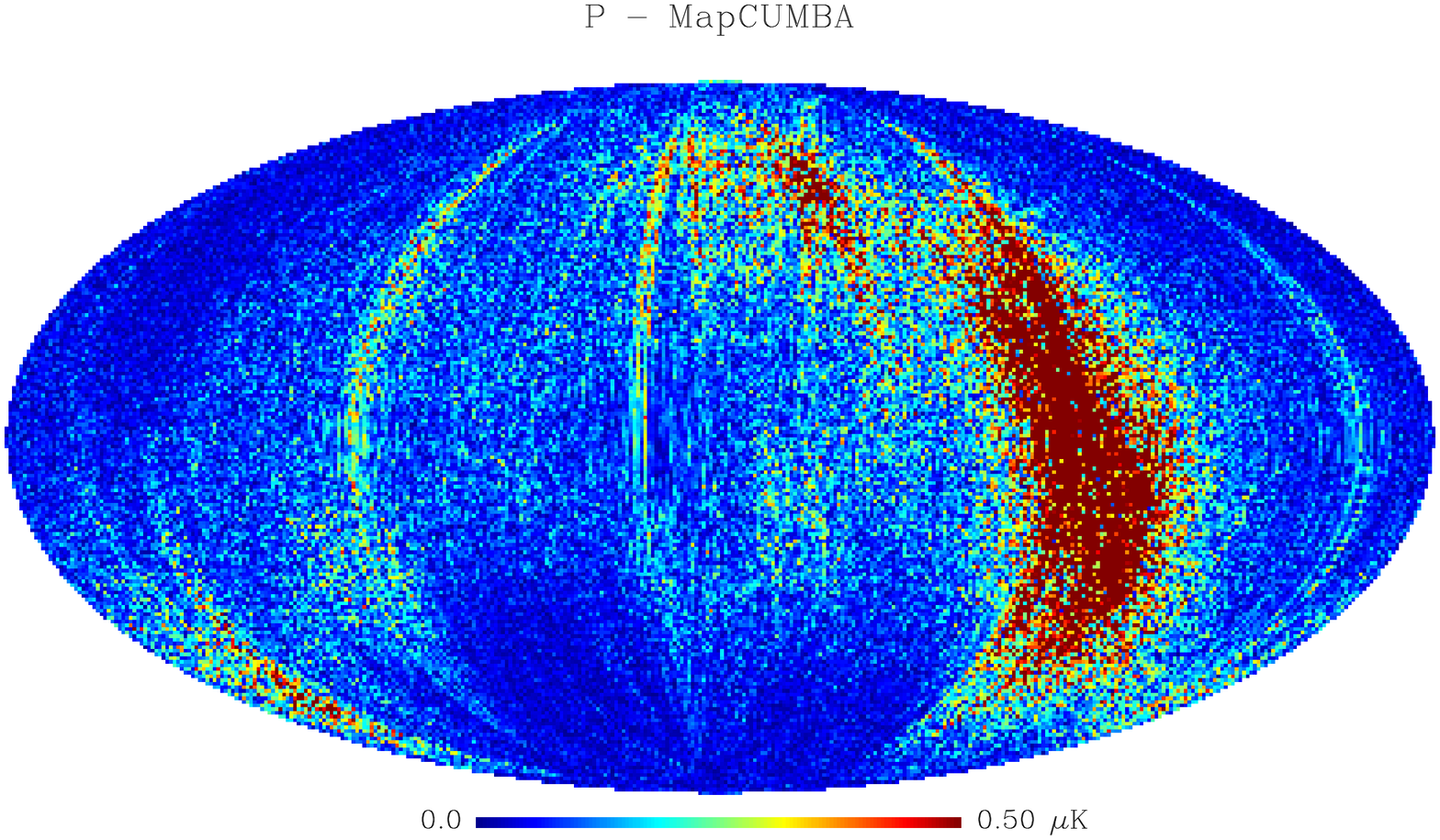}
    \includegraphics[scale=0.33,angle=90]{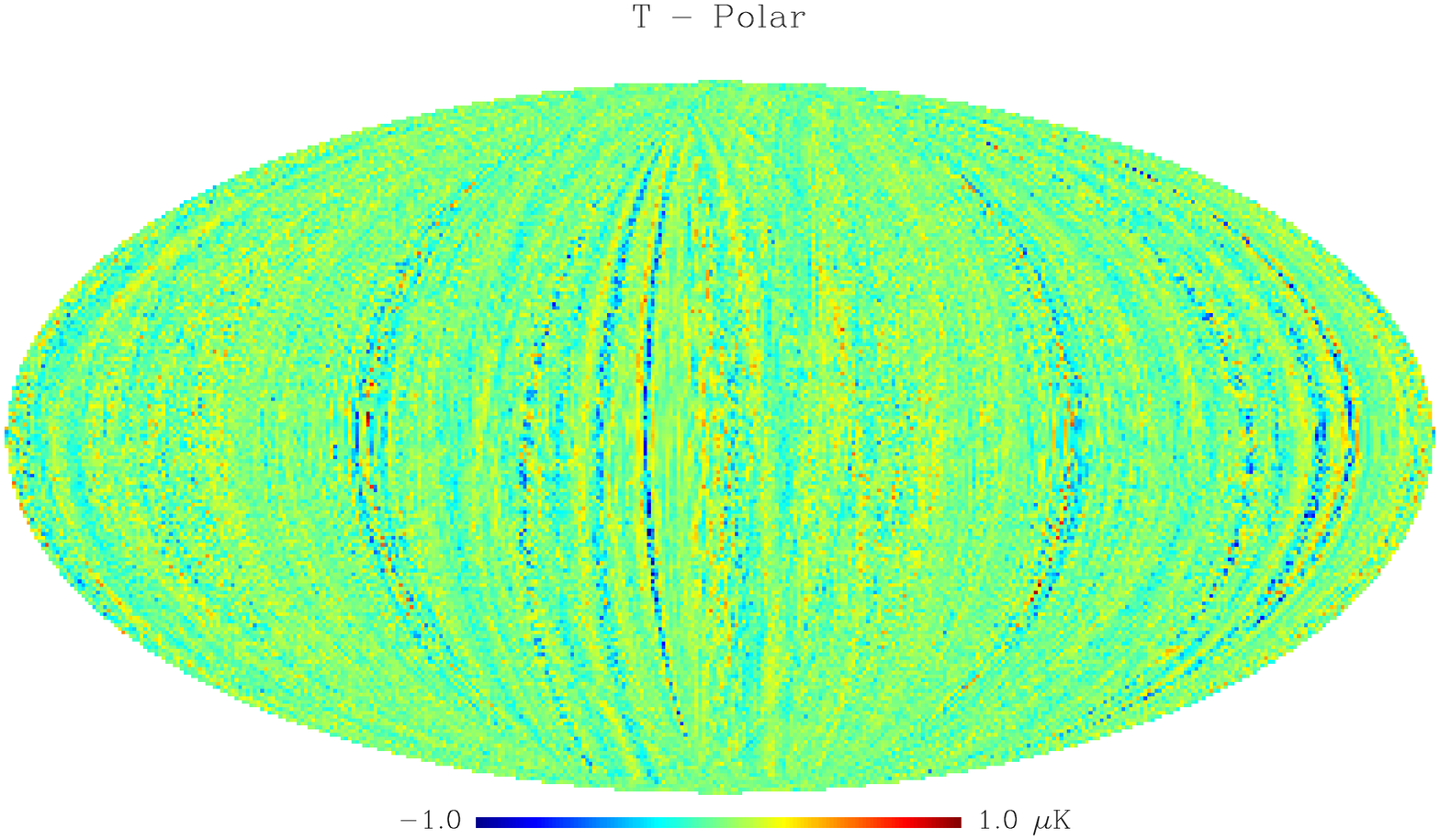}
    \includegraphics[scale=0.33,angle=90]{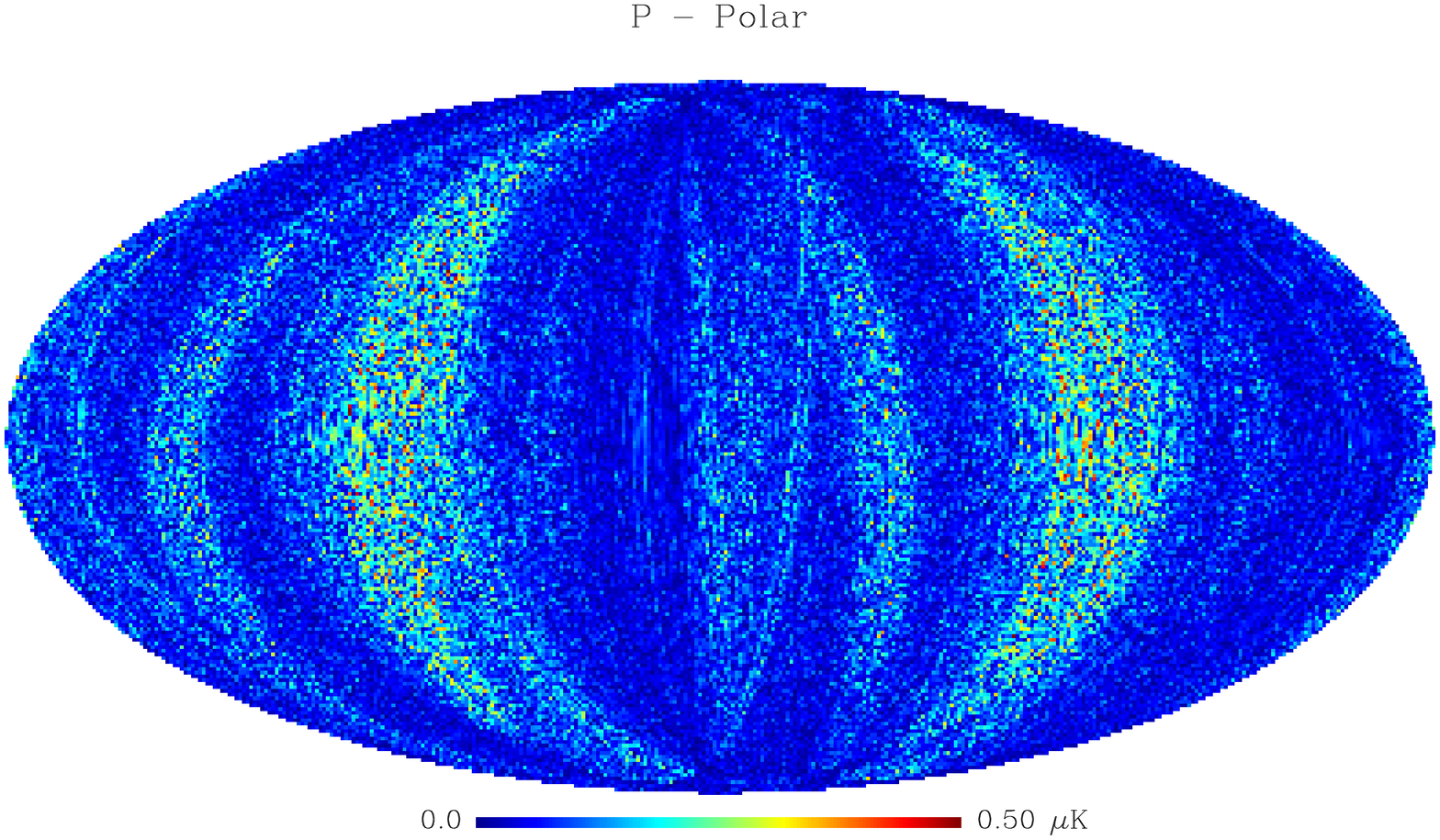}
    \end{center}
\caption{The difference maps between the noiseless output maps and
the binned noiseless maps for MapCUMBA (top row) and Polar (bottom
row). The maps contain CMB, dipole and foreground (CDF) and their
resolution is $N_{\rm side} = 512$. They are displayed here in
ecliptic coordinates. The left hand maps are for the Stokes I and
the right hand maps are for the magnitude $P$ of the polarisation
vector ($P = \sqrt{Q^2 + U^2}$). The map units are CMB microkelvins.
 (See www.helsinki.fi/\~{}tfo\_cosm/tfo\_planck.html for
better-quality figures.)}
  \label{fig:cdf_cdfbin}
 \end{figure}

\begin{figure} [!ht]
    \begin{center}
    \includegraphics[scale=0.33,angle=90]{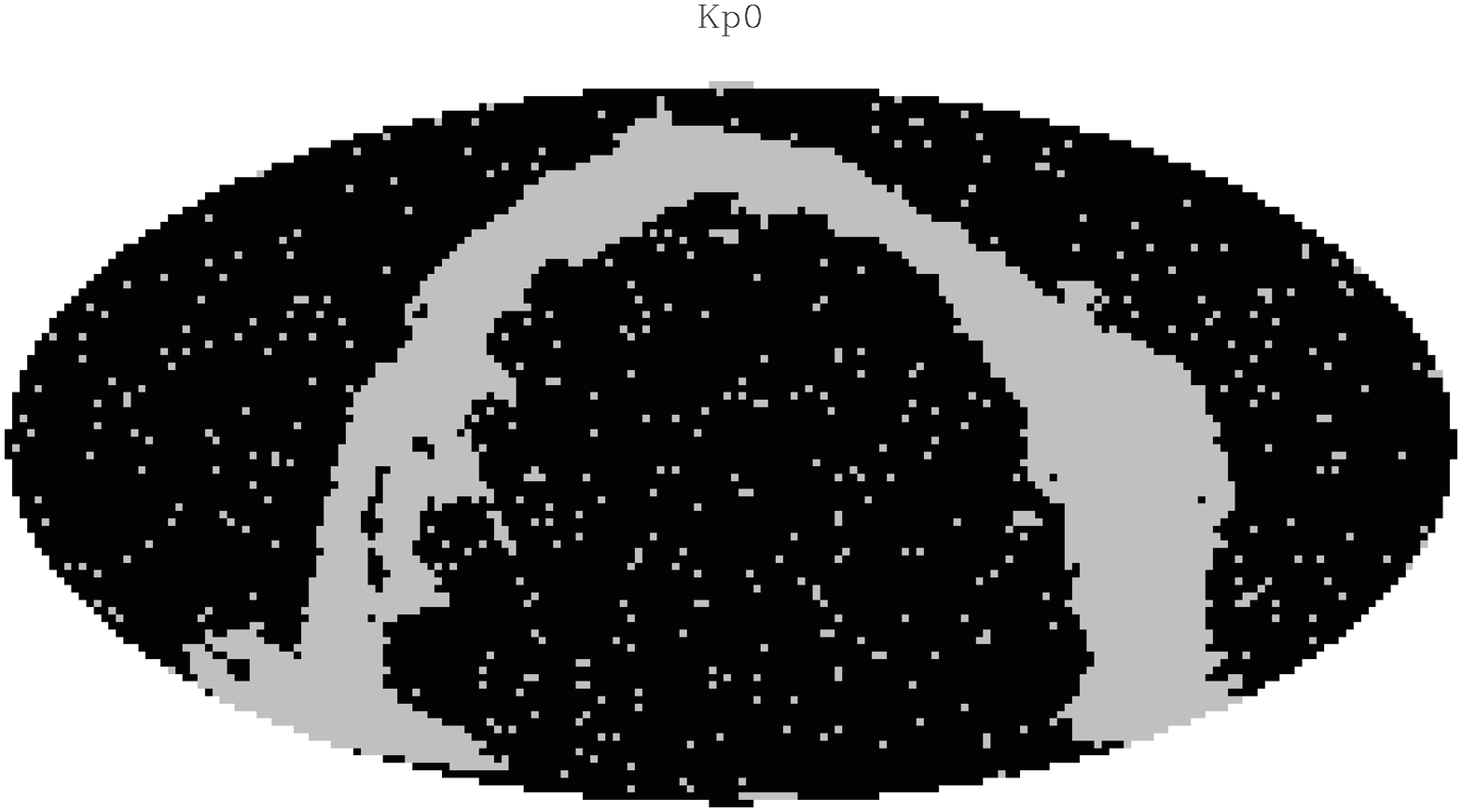}
    \includegraphics[scale=0.33,angle=90]{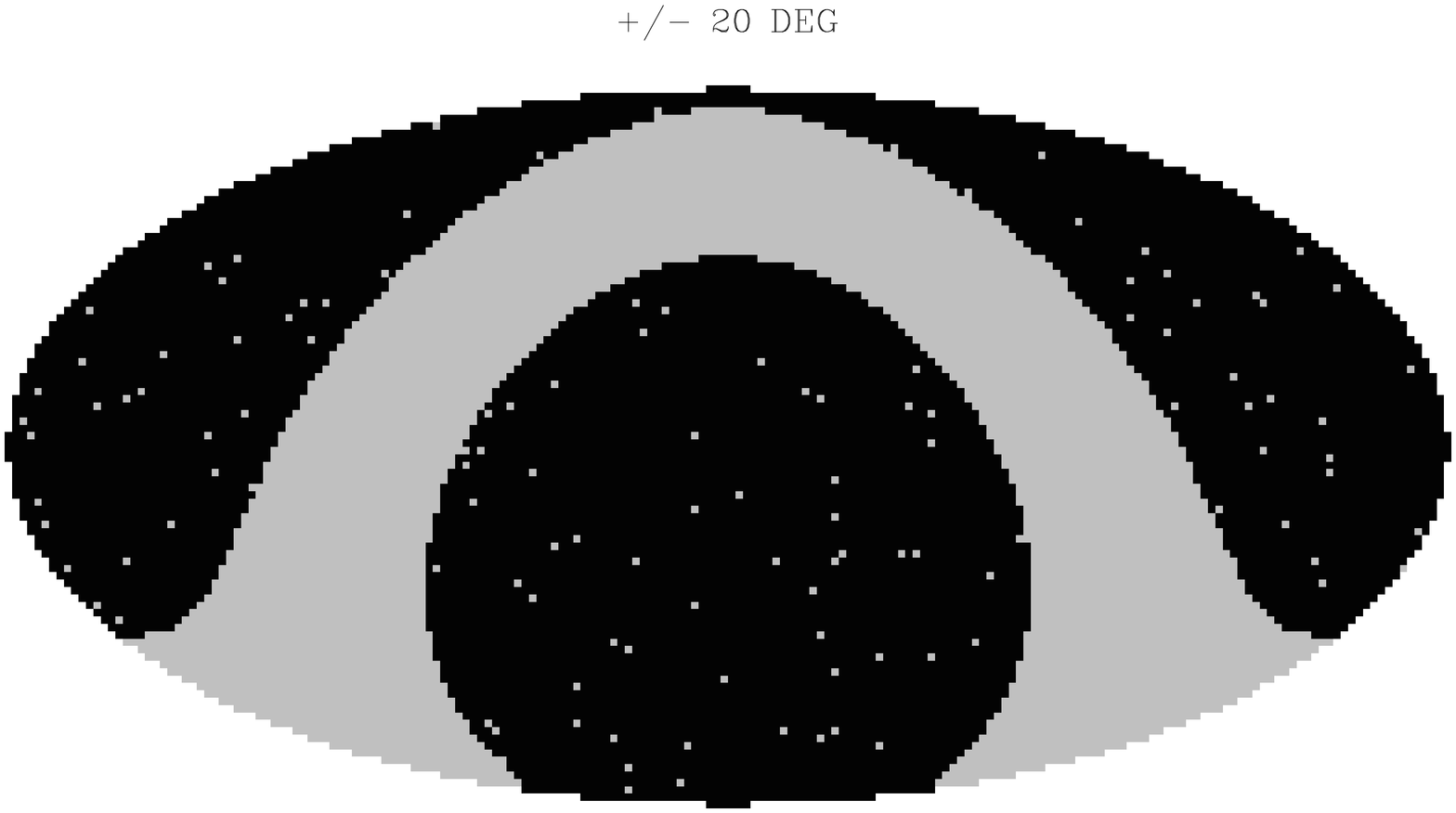}
    \end{center}
\caption{Masks showing two different schemes for removing the galaxy
and $\sim$2000 strongest extra-galactic point sources from our maps.
The light areas were removed. The left hand mask is the WMAP Kp0
mask (Bennett et al.~\cite{bennett03b}) after we had removed our
point sources from it.  We call this mask the Kp0 cut. The right
hand mask removes our point sources and a region, that is 20$\degr$
above and below the galactic plane. We call this mask the
$\pm$20$\degr$ cut. Note that the original WMAP Kp0 mask removes the
point sources identified by the WMAP project. We did not have the
WMAP point sources in our simulations, but we inserted our point
sources randomly (see Sect.~\ref{subsubsec:egs}). Therefore our Kp0
cut removes more point sources than the $\pm$20$\degr$ cut. The sky
coverage fractions of the masks are 0.758 (Kp0 cut) and 0.649
($\pm$20$\degr$ cut). (See
www.helsinki.fi/\~{}tfo\_cosm/tfo\_planck.html for better-quality
figures.)}
  \label{fig:masks}
 \end{figure}

To see what the magnitude of the signal error is outside the
galactic and point source regions, we removed the galaxy and the
strongest point sources from the output and binned noiseless maps
and examined the differences of these cut maps. We used the Kp0 and
$\pm$20$\degr$ cuts to remove the pixels (see Fig.~\ref{fig:masks}).
We recomputed the pixel statistics for the cut maps ($N_{\rm side} =
512$). The results are shown in Table~\ref{tab:cdf_512_cut}. Because
the erroneous pixels of the Polar and Springtide output maps are not
strongly localized in the galactic or point source regions (see the
bottom row of Fig.~\ref{fig:cdf_cdfbin}), the removal of these
pixels does not reduce the map errors significantly (see
Table~\ref{tab:cdf_512_cut}).  The errors of the GLS and Madam
output maps are reduced more, but they still remain larger than the
errors of the Polar and Springtide maps.

\begin{table}[h!]
\begin{center}
\begin{tabular}{| r  | c | c | c | c | c | c | c | c | c | c | c | c |}
\hline
 $N_{\rm side} = 512$ & \multicolumn{6}{c|}{CDF - CDFbin, rms $[\mu$K$]$} & \multicolumn{6}{c|}{max - min $[\mu$K$]$}\\
Code &  \multicolumn{2}{c|}{I}  & \multicolumn{2}{c|}{Q} & \multicolumn{2}{c|}{U} & \multicolumn{2}{c|}{I}  & \multicolumn{2}{c|}{Q} & \multicolumn{2}{c|}{U}  \\
\hline
Polar &  $0.144$ & $0.145$ & $0.0680$ & $0.0664$ & $0.0697$ & $0.0668$ & $2.46$ & $2.18$ & $1.20$ & $1.10$ & $1.28$ & $1.28$\\
Springtide &  $0.144$ &  $0.145$ & $0.0677$ & $0.0662$ & $0.0693$ & $0.0664$ & $2.38$ & $2.11$ & $1.19$ & $1.98$ & $1.27$ & $1.26$\\
Madam &  $0.239$ & $0.222$ & $0.102$ & $0.0823$ & $0.105$ & $0.0831$ & $8.97$ & $7.58$ & $3.31$ & $2.01$ & $4.27$ & $2.21$\\
MapCUMBA &  $0.247$ & $0.229$ & $0.103$ & $0.0835$ & $0.107$ & $0.0845$ & $9.96$ & $8.30$ & $3.38$ & $2.11$ & $4.35$ & $2.24$\\
MADmap &  $0.246$ & $0.229$ & $0.104$ & $0.0835$ & $0.107$ & $0.0846$ & $9.96$ & $8.30$ & $3.38$ & $2.11$ & $4.35$ & $2.24$\\
ROMA &  $0.246$ & $0.229$ & $0.104$ & $0.0835$ & $0.107$ & $0.0846$ & $9.96$ & $8.30$ & $3.38$ & $2.11$ & $4.35$ & $2.24$\\
\hline
  & Kp0 & $\pm$20$\degr$ & Kp0 & $\pm$20$\degr$ & Kp0 & $\pm$20$\degr$ & Kp0 & $\pm$20$\degr$ & Kp0 & $\pm$20$\degr$ & Kp0 & $\pm$20$\degr$\\
\hline
\end{tabular}
\end{center}
\caption{Same as Table~\ref{tab:cdf_512}, but now the galaxy and the
strongest extra-galactic point sources were removed from the maps
before computing the pixel statistics.  We applied two different
schemes for removing the pixels. Their masks are shown in
Fig.~\ref{fig:masks}. The data for the Kp0 and $\pm$20$\degr$ cuts
are shown in their own columns. The units are CMB microkelvins.}
\label{tab:cdf_512_cut}
\end{table}

\subsection{Sampling of the polarisation directions} \label{subsec:pol_directions}

We examined the effect of the {\it rcond\/} threshold on the pixel
statistics of the Polar residual noise and signal error maps. We
used three different {\it rcond\/} thresholds when discarding poorly
sampled pixels. The resulting statistics for the $N_{\rm side} = 1024$ maps are shown
in Table~\ref{tab:polar_cdf_1024} (for the signal error map) and in
Table~\ref{tab:polar_n_1024} (for the residual noise map). The {\it rcond\/} threshold
(i) (see Table~\ref{tab:polar_cdf_1024}) is the same as the smallest {\it rcond\/}
value of the pixels of our $N_{\rm side} = 512$ output maps.
Tables~\ref{tab:polar_cdf_1024} and~\ref{tab:polar_n_1024} show that the errors of
the output maps (especially in the polarisation maps) increase rapidly if we accept
poorly sampled pixels.

\begin{table}[h!]
\begin{tabular}{| r  | c c c | c c c |}
\hline
 $N_{\rm side} = 1024$   & \multicolumn{3}{c|}{CDF - CDFbin, rms $[\mu$K$]$} & \multicolumn{3}{c|}{max - min $[\mu$K$]$}\\
Code &   I   &   Q   &   U  & I   &   Q   &   U  \\
\hline
Polar ({\it rcond} $\geq$ 0.2165)  &  $0.115$ & $0.0642$ & $0.0654$ & $3.94$ & $1.48$ & $1.56$\\
Polar ({\it rcond} $\geq$ $10^{-2}$) &  $0.118$ & $0.0717$ & $0.0732$ & $4.63$ & $5.11$ & $5.33$\\
Polar ({\it rcond} $\geq$ $10^{-6}$)  &  $0.119$ & $0.4244$ & $0.4306$ & $4.76$ & $282.3$ & $289.4$ \\
\hline
\end{tabular}
\caption{Pixel statistics of the Polar noiseless difference maps: output map $-$
binned noiseless map.  Maps contained CMB, dipole, and foreground (CDF) and their
resolution was $N_{\rm side} = 1024$. Three different {\it rcond\/} thresholds were
applied when poorly sampled pixels were discarded before computing the statistics. The
thresholds and the corresponding numbers of discarded pixels (in parenthesis) were:
(i) {\it rcond\/} = 0.2165 (2,557,265), (ii) {\it rcond\/} = $10^{-2}$ (912,968), and
(iii) {\it rcond\/} = $10^{-6}$ (811,419). The pixels with no observations are included
in the discarded pixels. The {\it rcond\/} threshold (i) is the same as the smallest
{\it rcond\/} of the pixels of our $N_{\rm side} = 512$ output maps. The units are CMB
microkelvins.} \label{tab:polar_cdf_1024}
\end{table}

\begin{table}[h!]
\begin{tabular}{| r  | c c c | c c c |}
\hline
 $N_{\rm side} = 1024$   & \multicolumn{3}{c|}{Res. noise, rms $[\mu$K$]$} & \multicolumn{3}{c|}{max - min $[\mu$K$]$}\\
Code &   I   &   Q   &   U  & I   &   Q   &   U  \\
\hline
Polar ({\it rcond} $\geq$ 0.2165)  &  $86.33$ & $126.60$ & $127.60$ & $4669$ & $6763$ & $6335$\\
Polar ({\it rcond} $\geq$ $10^{-2}$) &  $97.83$ & $196.84$ & $198.90$ & $4669$ & $20422$ & $19315$\\
Polar ({\it rcond} $\geq$ $10^{-6}$)  &  $99.74$ & $1564.0$ & $1563.8$ & $9338$ & $1222731$ & $1201932$ \\
\hline
White noise ({\it rcond} $\geq$ 0.2165)  &  $85.52$ & $125.50$ & $126.50$ &  &  & \\
White noise ({\it rcond} $\geq$ $10^{-2}$) &  $96.93$ & $195.18$ & $197.13$ & & & \\
White noise ({\it rcond} $\geq$ $10^{-6}$)  &  $98.86$ & $1608.8$ & $1596.0$ & & & \\
\hline
\end{tabular}
\caption{Same as Table~\ref{tab:polar_cdf_1024}, but the statistics
are computed for the Polar residual noise maps. The map resolution
was $N_{\rm side} = 1024$. The actual number of hits and the actual
sampling of the polarisation directions were considered for every
non-discarded pixel when the approximate rms of the white noise
maps were computed.} \label{tab:polar_n_1024}
\end{table}

\section{Conclusions}

We compared the output maps of three GLS map-making codes (MapCUMBA,
MADmap, and ROMA) and three destriping codes (Polar, Springtide, and
Madam). We made maps from simulated observations of four LFI 30~GHz
detectors of the {\sc Planck} satellite. The observed signal was
comprised of dipole, CMB, diffuse galactic emissions, extragalactic
radio sources, and detector noise. CMB, galactic and radio source
signals included both total intensity (temperature) and
polarisation. We assumed identical circularly symmetric beams for
every detector. The GLS codes and Madam require information on the
detector noise. Polar and Springtide do not rely on that
information. In this study we assumed that the detector noise
spectra were perfectly known.

We subtracted the binned noiseless map from the output maps of the
map-making codes and examined the remaining residual map. It is a
sum of two components: residual noise (due to detector noise) and
signal error that arises from the subpixel signal structure that
couples to the output map through the map-making.

The maps of the GLS codes have nearly the same residual noise
levels, which are lower than the noise of the Polar and Springtide
maps.  Madam can produce maps with as low noise as the GLS codes.
The residual noise of Springtide is higher than the noise of Polar,
because Springtide works with long (1~hr) ring baselines rather than
with shorter baselines used in Polar (1~min). These differences in
the residual noise between the codes are, however, small.

The signal error was smallest in the Springtide maps, but the signal
error of Polar was just a bit larger. The signal errors of the GLS
codes were more significant. The signal error in the Madam maps was
slightly smaller than in the GLS maps. However, the signal errors of
all map-making codes were significantly smaller than their residual
$1/f$ noise.

We examined several schemes to reduce the signal error. In
destriping the most effective method was the reduction of the pixel
size of the crossing points. For example, for $N_{\rm side} = 512$
output maps we could reduce the rms of the signal error by a factor
$\backsim$2 by using $N_{\rm side} = 1024$ pixels (instead of
$N_{\rm side} = 512$ pixels) for the crossing points. We did not
find methods to bring similar improvement in the GLS signal error.
Reducing the crossing point pixel size in destriping increases the
map noise slightly.

The next step in the map-making studies of the {\sc Planck} CTP
working group is to change to asymmetric beams and assume
integration over non-zero sample intervals as detectors scan across
the sky. This work has started and we will make maps from the
simulated LFI 30~GHz observations using elliptic Gaussian beams that
are fits to the realistic beam simulations. We will examine e.g. the
quality of the maps and the magnitude of the leakage from
temperature to polarisation due to differences between the responses
of the detector beams.

\begin{acknowledgements}
  The work reported in this paper was done by the CTP Working Group of
  the {\sc Planck} Consortia. {\sc Planck} is a mission of the
  European Space Agency.
  The authors would like to thank Institut d'Astrophysique de Paris (IAP) for its
  hospitality in June 2005 when the CTP Working Group met to undertake
  this work.
  This research used resources of the National Energy Research
  Scientific Computing Center, which is supported by the Office of
  Science of the U.S. Department of Energy under Contract
  No. DE-AC03-76SF00098.
  We acknowledge the use of version 0.1 of the Planck reference sky model,
  prepared by the members of the Planck Working Group 2 and available at
  http://www.planck.fr/heading79.html.
  This work has made use of the {\sc Planck} satellite simulation
  package (Level S), which is assembled by the Max Planck Institute
  for Astrophysics {\sc Planck} Analysis Centre (MPAC).
  EK and TP were supported by the Academy of Finland grants no.
  205800, 213984, and 214598. They also thank the von Frenckell
  foundation for financial support.
  Some of the results in this paper have been derived using the
  HEALPix package (G\'orski et al.~\cite{gorski99}, \cite{gorski05a}).
  The US {\sc Planck} Project is supported by the NASA
  Science Mission Directorate.
\end{acknowledgements}


\begin{thebibliography}{}

\bibitem[2006]{ashdown06}
  Ashdown, M. 2006, submitted to A\&A, [astro-ph/0606348]

\bibitem[2003]{baccigalupi03}
Baccigalupi C., 2003, New Astron. Rev., 47, 1127

\bibitem[2003]{bennett03b}
Bennett C.L. et al., 2003, ApJS, 148, 97

\bibitem[Beno\^{\i}t et al. 2003]{benoit03}
Beno\^{\i}t, A., Ade, P., Amblard, A. et al. 2003, A\&A, 399, L19

\bibitem[2004]{benoit04}
Beno\^{\i}t A. et al., 2004, A\&A, 424, 571

\bibitem[Carretti et al. 2005]{carretti05}
Carretti E., Bernardi G., Sault R.J., Cortiglioni S., Poppi S.,
2005, MNRAS, 358, 1

\bibitem[2000]{challinor00}
 Challinor, A., Fosalba, P., Mortlock, D., et al. 2000, Phys.Rev.D,
 62, 123002

\bibitem[2001]{dore01} Dor\'e, O., Teyssier, R., Bouchet,
  F.R., Vibert, D. \& Prunet, S., 2001, \aap, 374, 358; see also
   http://ulysse.iap.fr/cmbsoft/mapcumba/

\bibitem[1999]{duncan99} Duncan, A.R., Reich, P., Reich, W.,
F\"urst, E. 1999, A\&A, 350, 447

\bibitem[2005]{dupac05}
 Dupac, X., \& Tauber, J. 2005, A\&A, 430, 363

\bibitem[2005]{efstathiou05} Efstathiou, G., Lawrence, C., Tauber, J. et al. 2005,
"{\sc Planck} - The Scientific Programme", ESA-SCI(2005)1, "{\sc
Planck} Bluebook" available in http://www.rssd.esa.int/Planck.

\bibitem[Finkbeiner et al. 1999]{finkbeiner99}
Finkbeiner D.P., Davis M., Schlegel D.J. 1999, ApJ, 524, 867

\bibitem[Giardino et al. 2002]{giardino02}
Giardino G. et al., 2002, A\&A, 387, 82

\bibitem[1999]{gorski99} G\'orski, K.M., Hivon, E., \& Wandelt, B.D. 1999,
 in Proceedings of the MPA/ESO Cosmology Conference "Evolution of
 Large-Scale Structure", ed. A.J. Banday, R.S. Sheth, \& L. Da Costa,
 PrintPartners Ipskamp, NL, 37, [astro-ph/9812350]

\bibitem[2005a]{gorski05a} G\'orski, K.M., Hivon, E., Banday, A.J., et al. 2005, ApJ, 622, 759

\bibitem[2005b]{gorski05b} G\'orski, K.M., Wandelt, B.D., Hivon, E.,
 Hansen, F.K., \& Banday, A.J. 2005,
 "The HEALPix Primer" (Version 2.00), available
 in http://healpix.jpl.nasa.gov

\bibitem[Haslam et al. 1982]{haslam82}
Haslam C.G.T., Stoffel H., Salter C.J., Wilson W.E., 1982, A\&AS,
47, 1

\bibitem[2006]{hivon06} Hivon, E. et al. 2006, under preparation

\bibitem[2006]{jarosik06} Jarosik, N. et al. 2006, submitted

\bibitem[2006]{larquere06} Larqu\`{e}re, L. 2006, Ph.D. Thesis,
Universite Paris 7 Denis-Diderot

\bibitem[2004]{odwyer04} O'Dwyer, I.J. et al. 2004, ApJ, 617, L99

\bibitem[2006]{poutanen06}
 Poutanen, T., de Gasperis, G., Hivon, E., et al. 2006, A\&A, 449, 1311

\bibitem[Prunet et al.~1998]{prunet98}
Prunet S., Sethi S.K., Bouchet F.R., Miville-Deschenes M.A., 1998,
A\&A, 339, 187

\bibitem[Reich \& Reich~1986]{Reich}
  Reich, P. \& Reich, W. 1986, A\&A Suppl, 63, 205

\bibitem[2006]{reinecke06}
  Reinecke, M., Dolag, K., Hell, R., Bartelmann, M., \& En\ss lin, T. 2006, A\&A,
  445, 373

\bibitem[Tucci et al.~2002]{tucci02}
Tucci M., et al., 2002, ApJ, 579, 607

\bibitem[Tucci et al.~2004]{tucci04}
Tucci M., Martinez-Gonzalez E., Toffolatti L., Gonzalez-Nuevo J., De
Zotti G., 2004, MNRAS, 349, 1267

\bibitem[Uyaniker et al. 1999]{uyaniker99}
Uyaniker B., F$\ddot{\rm u}$rst E., Reich W., Reich P., Wielebinski
R., 1999, A\&AS, 138, 31

\bibitem[2001]{wandelt01}
  Wandelt, B. D., \& G\'orski, K. M. 2001, Phys.Rev.D, 63, 123002

\end{thebibliography}
\end{document}